\documentclass[a4paper,10pt]{article}

\usepackage{fullpage} 
\usepackage{parskip} 
\usepackage{tikz} 
\usepackage{amsmath}
\usepackage{hyperref}
\usepackage{graphicx}
\usepackage[font=footnotesize]{caption}
\usepackage{array}
\usepackage{multicol}
\usepackage{amssymb}
\usepackage[ruled,vlined,linesnumbered]{algorithm2e}
\usepackage{mathrsfs}
\usepackage{bm}
\SetAlFnt{\footnotesize}
\usepackage{adjustbox}
\graphicspath{ {images/} }
\usepackage{color}
\usepackage{hyperref}
\usepackage{booktabs}
\usepackage{multirow}
\usepackage[sort,numbers]{natbib}
\usepackage[boldmath]{numprint}
\usepackage{tikz}
\usetikzlibrary{bayesnet}
\usetikzlibrary{arrows}
\usepackage{subcaption}
\usepackage{subfiles}
\usepackage{mathtools}
\usepackage{breqn}
\usepackage{lscape}
\usepackage{rotating}

\usepackage{color}
\definecolor{darkgreen}{rgb}{0,0.6,0.2}

\begin{document}
\title{Deep Generative Models for \\ Reject Inference in Credit Scoring}
\author{Rogelio A. Mancisidor$^{a,b,\ast}$ \\ \href{mailto:rogelio.a.mancisidor@uit.no}{rogelio.a.mancisidor@uit.no}
\and Michael Kampffmeyer $^a$ \\
\href{mailto:michael.c.kampffmeyer@uit.no}{michael.c.kampffmeyer@uit.no}
\and Kjersti Aas $^c$ \\
\href{mailto:kjersti@nr.no}{kjersti@nr.no}
\and Robert Jenssen $^a$ \\
\href{mailto:robert.jenssen@uit.no}{robert.jenssen@uit.no}
\\
\\
\normalsize{$^{a}$Machine Learning Group, Department of Physics and Technology, Faculty of Science and}\\
\normalsize{Technology, UiT - The Arctic University of Norway, Hansine Hansens veg 18, Troms{\o} 9037, Norway}\\
\normalsize{$^b$Credit Risk Models, Santander Consumer Bank AS, Strandveien 18, Lysaker 1325, Norway}\\
\normalsize{$^{c}$Statistical Analysis, Machine Learning and Image Analysis}\\
\normalsize{Norwegian Computing Center, Gaustadalleen 23a, Oslo 0373, Norway}\\
\\
\normalsize{$^\ast$Corresponding author}
\
}
\maketitle

\renewenvironment{abstract}
{\begin{quote}
\noindent \rule{\linewidth}{.5pt}\par{\bfseries \abstractname.}}
{\noindent 
\rule{\linewidth}{.5pt}
\end{quote}
}

\begin{abstract}
Credit scoring models based on accepted applications may be biased and their consequences can have a statistical and economic impact. Reject inference is the process of attempting to infer the creditworthiness status of the rejected applications. Inspired by the promising results of semi-supervised deep generative models, this research develops two novel Bayesian models for reject inference in credit scoring combining Gaussian mixtures and auxiliary variables in a semi-supervised framework with generative models. To the best of our knowledge this is the first study coupling these concepts together. The goal is to improve the classification accuracy in credit scoring models by adding reject applications. Further, our proposed models infer the unknown creditworthiness of the rejected applications by exact enumeration of the two possible outcomes of the loan (default or non-default). The efficient stochastic gradient optimization technique used in deep generative models makes our models suitable for large data sets. Finally, the experiments in this research show that our proposed models perform better than classical and alternative machine learning models for reject inference in credit scoring, and that model performance increases with the amount of data used for model training. 


\end{abstract}

\section{Introduction}\label{intro}
Credit scoring uses statistical models to transform the customers' data into a measure of the borrowers' ability to repay the loan \cite{anderson2007credit}. These models are developed, commonly, based on accepted applications because the bank knows whether the customer repaid the loan. The problem is that this data sample is biased since it excludes the rejected applications systematically. This is called selection bias.

Using a biased sample to estimate any model has several problems. The straightforward consequence is that the model parameters are biased \cite{bucker2013reject}, which has a statistical and economic impact \cite{nguyen2016,chen2001economic}. Another consequence is that the default probability can be underestimated, affecting the risk premium and the profitability of the bank \cite{marshall2010variable}. Hence, reject inference, which is the process of attempting to infer the true creditworthiness status of the rejected applications \cite{hand1993can}, has created a great deal of interest.

There is a vast literature on reject inference using classical statistical methods. However, there has been little research using machine learning techniques (see Table \ref{tbl_RIoverview}). Semi-supervised learning approaches design and train models using labeled (accepted applications) and unlabeled data (rejected applications), and aim to utilize the information embedded in both data to improve the classification of unseen observations. There are several fields where semi-supervised deep generative models have achieved state-of-the-art results, e.g. in semi-supervised image classification \cite{kingma2014semi,maaloe2016auxiliary}, in semi-supervised sentiment analysis \cite{wu2019semi,fu2019semi}, and in unsupervised clustering \cite{zheng2016variational}. Additionally, the useful information embedded in their latent space is well documented \citep{bowman2015generating,hou2017deep,latif2017variational,mancisidor2018segment}. Inspired by the modeling framework introduced by \cite{kingma2014semi}, this research develops two novel models for reject inference models in credit scoring combining, for the first time, auxiliary variables \citep{maaloe2016auxiliary} and Gaussian mixtures parametrized by neural networks in a semi-supervised framework.

Our proposed models have a flexible latent space, induced by the Gaussian mixtures, to improve the variational approximation and the reconstruction of the input data \cite{maaloe2015improving,maaloe2016auxiliary}. In addition, one of our models not only uses the input data to classify new loan applications, but also a latent representation of it. This makes the classifier more expressive \cite{maaloe2015improving,maaloe2016auxiliary}. We compare the performance of the semi-supervised generative models with a range of techniques representing the state-of-the-art in reject inference for credit scoring, including three classical reject inference techniques (reclassification, fuzzy parceling\footnote{For a review of the reclassification and fuzzy parceling approaches see \cite{anderson2007credit,nguyen2016}.} and augmentation \cite{hsia1978credit}), and three semi-supervised machine learning approaches (self-learning \cite{rosenberg2005semi} MLP, self-learning SVM, and semi-supervised SVM \cite{gieseke2012sparse}). Additionally, we include two supervised machine learning models (multilayer perceptron (MLP) \cite{rumelhart1985learning} and support vector machine (SVM) \cite{cortes1995support}) to measure the marginal gain of reject inference. 

To summarize, the main contributions of this paper are as follows:
\begin{enumerate}
    \item We develop two novel reject inference models for credit scoring combining auxiliary variables and Gaussian mixtures in a semi-supervised framework with generative models for the first time.
    
    \item We derive the objective functions for our proposed models and show how they can be parameterized by MLPs and optimized with stochastic gradient descent.
    
    \item We parametrize the Gaussian mixtures using an MLP and we show how to train them with semi-supervised data.
    
    \item Our empirical results show that our proposed models achieve higher performance compared to the state-of-art methods in credit scoring. Additionally, the model performance for our proposed models increases with the amount of data used for training.
\end{enumerate}

The rest of the paper is organized as follows. Section \ref{sec_relatedwork} reviews the related work on reject inference in credit risk, then Section \ref{sec_ssl} presents an overview of semi-supervised deep generative models and introduces the proposed models. Section \ref{sec_results} explains the data, methodology and main results. Finally, Section \ref{sec_conclusion} presents the main conclusion of this research.

\section{Related Work}\label{sec_relatedwork}
Banks decide whether to grant credit to new applications as well as how to deal with existing customers, e.g. deciding whether credit limits should be increased and determining which marketing campaign is most appropriate. The tools that help banks with the first problem are called credit scoring models, while behavioral scoring models are used to handle exiting customers \cite{thomas2000survey}. Both type of models estimate the ability that a borrower will be unable to meet its debt obligations, which is referred to as default probability. This research focuses on reject inference to improve the classification accuracy of credit scoring models by utilizing the rejected applications. In Table (\ref{tbl_RIoverview}), we present an updated research overview on reject inference in credit scoring extending the one presented in \cite{li2017reject}.

There are two broad approaches to estimate the default probability; the function estimation model (e.g. logistic regression) and the density estimation approach (e.g. linear discriminant analysis). The latter is more susceptible to provide biased parameter estimates when the rejected applications are ignored \cite{feelders2000credit,hand1993can}.

According to \cite{hand1993can}, reject inference represents several challenges. First of all, when attempting to correct the selection bias, the customer characteristics used to develop the current credit scoring model must be available. Otherwise, including the rejected applications in the new model might be insufficient to correct the selection bias. Some techniques, such as mixture decomposition, require assumptions about the default and non-default distributions. In general, these distributions are unknown. Finally, the methods based on supplementary credit information about the reject applications, which might be bought at credit bureaus, can be unrealistic for some financial institutions. Either they cannot afford to pay for it or the data may not be available.

\begin{table}[t!]
\centering 
\begin{adjustbox}{width=\textwidth}
\begin{tabular}{|l|l|l|c|c|l|l|}
\hline
(Year) \  Author & Data type & Status of rejects & No. of accepts & No. of rejects & Reject Inference approach  & Classification method \\
\hline
(1993) \citet{joanes1993reject}      & Artificial    &Unknown    &75 &12 &Reclassification   &Logistic\\
(2000) \citet{feelders2000credit}    & Artificial&Unknown    &Varying    &Varying    &EM & QDA, Logistic\\
(2001) \citet{chen2001economic}  & Coorporate   & Known  & 298  & 599  & Heckman's model  & Probit, Bivariate probit\\
(2003)  \citet{banasik2003sample} & Consumer   &Known  &8 168  &4 040  &Augmentation   &Logistic,  Probit\\
(2004) \citet{crook2004does}    &Consumer   &Known  &8 168  &4 040  &Augmentation, Extrapolation &Logistic\\
(2004) \citet{verstraeten2005impact} & Consumer   &Partially known    &38 048 &6 306 &Augmentation   &Logistic\\
(2005) \citet{banasik2005credit}    & Consumer   &Known  &8 168  &4 040  &Augmentation   &Logistic\\
(2006) \citet{sohn2006reject}*    &Consumer   &Unknown    &759    &10 &Reclassiﬁcation    &Survival analysis\\
(2007) \citet{banasik2007reject}   &Consumer   &Known  &8 168  &4 040  &Augmentation and Heckman's model   &Logistic, Bivariate probit\\
(2007) \citet{kim2007technology} & Corporate  &Known  &4 298  &689    & Heckman's model  & Bivariate  probit\\
(2007) \citet{wu2007handling} & Artificial  & Known  & Varying  & Varying    & Heckman's model  & OLS, Bivariate  Probit\\
(2010) \citet{banasik2010reject}*    &Consumer   &Known  &147 179    &Varying    &Augmentation   &Survival analysis\\
(2010)  \citet{marshall2010variable}  & Consumer   & Known  & 40 700    & 2 934    & Heckman's model   & Probit, Bivariate probit\\
(2010) \citet{maldonado2010semi} & Consumer   &Known  &800    &200    &Extrapolation  &SVM\\
(2012) \citet{chen2012bound} &Corporate  &Known  &4 589  &Varying    &Bound and Collapse &Bayesian\\
(2013) \citet{bucker2013reject}    &Consumer   &Unknown    &3 984  &5 667  &Augmentation    &Logistic\\
(2013) \citet{anderson2013modified} &Consumer   &Unknown    &3 000  &1 500  &Augmentation, EM   &Logistic\\
(2016) \citet{nguyen2016} & Consumer   & Unknown    &  56 016  & 142 571  & Augmentation, Extrapolation   & Logistic\\
(2017) \citet{li2017reject} & Consumer  & Unknown   & 56 626    & 563 215   & Extrapolation     & Semi-supervised SVM\\
\hline
\end{tabular}
\end{adjustbox}
\caption{Up to date research overview on reject inference. The scope of the research marked with * differs from ours, hence they are included in Section \ref{sec_relatedwork}.}
\label{tbl_RIoverview}
\end{table}

A simple approach for reject inference is augmentation \cite{hsia1978credit}. In this approach, the accepted applications are re-weighted to represent the entire population. The common way to find these weights is using the accept/reject probability. For example if a given application has a probability of being rejected of 0.80, then all similar applications would be weighted up $1/(1-0.8)=5$ times \cite{anderson2007credit}. None of the empirical research using augmentation shows significant improvements in either correcting the selection bias or improving model performance, see  \cite{anderson2007credit,banasik2005credit,banasik2007reject,banasik2003sample,bucker2013reject,crook2004does,verstraeten2005impact}. The augmentation technique assumes that the default probability is independent of whether the loan is accepted or rejected \cite{ash2002best}. However, \cite{kim2007technology} shows empirically that this assumption is wrong.

Heckman's bivariate two-stage model \cite{heckman1976common,heckman1979sample} has been used in different reject inference studies\footnote{The Heckman's model, named after Nobel Laureate James Joseph Heckman, has been extended or modified in different directions. See \cite{chen2001economic} for a chronological overview of the model evolution and its early applications. It was in \cite{boyes1989econometric} where the Heckman's approach was first applied to credit scoring where the outcome is discrete.}. This approach simultaneously models the accept/reject and default/non-default mechanisms. Assuming that the error terms in these processes are bivariate normally distributed with unit variance and correlation coefficient $\rho$, the selection bias arises when $\rho \neq 0$ and it is corrected using the inverse of the Mills ratio. 

Despite the popularity of Heckman's model, it is unclear whether this model can correct the selection bias or improve model performance. Some studies claim either higher model performance or different model parameters after using Heckman's model \cite{greene1998sample,banasik2003sample,banasik2007reject,kim2007technology,marshall2010variable}. These results, as explained by \cite{chen2001economic}, depend upon whether the selection and default equations are correlated. On the other hand, \cite{puhani2000heckman,wu2007handling,chen2012bound} state that the model parameters are inefficient, and the main criticism is that the Heckman's model fails to correct the selection bias when it is strong. This happens either when the correlation between the error terms in the selection and outcome equations is high or the data has high degree of censoring \cite{puhani2000heckman}. 

A comparison of different reject inference methods, e.g. augmentation, parceling, fuzzy parceling and the Heckman's model, is presented in \cite{nguyen2016}. The parceling and fuzzy parceling methods are very similar. They first fit a logistic regression model using the accepted applications. Then they use this model to estimate the default probability for all rejected applications. The difference is that the parceling method chooses a threshold on the default probability to assign the unknown outcome $y$ to the rejected applications. On the other hand, the fuzzy parceling method assumes that each reject application has both outcomes $y=1$ and $y=0$, with weights given by the fitted model using only the accepted applications. Finally, the parcelling (fuzzy parceling) method fits a new (weighted) logistic regression using both accepted and rejected applications. The results in \cite{nguyen2016} do not show higher model performance using the reject inference methods. However, the parameter estimates are different when applying the augmentation and parceling approaches. Hence, reject inference has a statistical and economic impact on the final model in this case.

Support vector machines are used in \cite{maldonado2010semi} to extend the self-training (SL) algorithm, by adding the hypothesis that the rejected applications are riskier\footnote{The self-training algorithm is an iterative approach where highly confident predictions about the unlabeled data are added to retrain the model. This procedure is repeated as many times as the user specify it. The main criticism of this method is that it can strengthen poor predictions \cite{kingma2014semi}.}. Specifically, their approach iteratively adds rejected applications with higher confidence, i.e. vectors far from the decision-hyperplane, to retrain a SVM (just as in the SL algorithm). However, vectors close to the hyperplane are penalized since the uncertainty about their true label is higher. Their proposed iterative approach shows superior performance compared to other reject inference configurations using SVMs, including semi-supervised support vector machines (S3VM). In addition to higher performance, the iterative procedure in \cite{maldonado2010semi} is faster than the S3VM.

The S3VM model is used in \cite{li2017reject} for reject inference in credit scoring \footnote{The model used in \cite{li2017reject}, originally developed by \cite{TianL17}, uses a branch-and-bound approach to solve the mixed integer constrained quadratic programming problem faced in semi-supervised SVMs. This approach reduces the training time making it suitable for large-sized problems.} using the accepted and rejected applications to fit an optimal hyperplane with maximum margin. The hyperplane traverses trough non-density regions of rejected applications and, at the same time, separates the accepted applications. Their results show higher performance compared to the logit and supervised support vector machine models. In Section \ref{sec_results}, we show that S3VM does not scale to large credit scoring data sets and that our proposed models are able to use, at least, 16 times more data compared to S3VM.

In \cite{feelders2000credit} Gaussian mixture models (GMM) are used for density estimation of the default probability. The idea is that each component in the mixture density models a class-conditional distribution. Then, the model parameters are estimated using the expectation-maximization (EM) algorithm, which can estimate the parameters even when the class labels for the rejected applications are missing. The EM algorithm is also used for reject inference in \cite{anderson2013modified}. Both papers report high model performance. However, the results in \cite{feelders2000credit} are based on artificial data and \cite{anderson2013modified} only judge performance based on the Confusion matrix. Finally, the major limitation of the EM algorithm is that we need to be able to estimate the expectation over the latent variables. We show in Section \ref{sec_ssl} that deep generative models circumvent this restriction by approximation.

A Bayesian approach for reject inference is presented in \cite{chen2012bound}. In this method the default probability is inferred from the missing data mechanism. The authors use the bound-collapse approach \footnote{This model is originally presented in Sebastiani and Ramoni (2000) "Bayesian inference with missing data using bound and collapse".} to estimate the posterior distribution over the score and class label, which is assumed to have a Dirichlet distribution as well as the marginal distribution of the missing class label. The reason for using the bound-collapse method is to avoid exhaustive numerical procedures, like the Gibbs Sampling, to estimate the posterior distributions in this model. Their results show that the Bayesian bound-collapse method perform better than the augmentation and Heckman's model.  

In this research we propose a novel Bayesian inference approach for reject inference in credit scoring, which uses Gaussian mixture models and differs from \cite{chen2012bound,feelders2000credit} in that our models are based on variational inference, neural networks, and stochastic gradient optimization. The main advantages of our proposed method are that (i) inference of the rejected applications is based on an approximation of the posterior distribution and on the exact enumeration of the two possible outcomes that the rejected applications could have taken, (ii) the models use a latent representation of the customers' data, which contain powerful information, and (iii) deep generative models scale to large data sets.  

\section{Deep Generative Models}\label{sec_ssl}
The principles of variational inference with deep neural networks are given in \cite{kingma2013auto,rezende2014stochastic}. Building upon this work, \cite{kingma2014semi} proposed a generalized probabilistic approach for semi-supervised learning. This approach will be explained in Section \ref{sec_ssdgm} before we introduce two novel models for reject inference in credit scoring in Sections \ref{sec_model1} and \ref{sec_model2}. 

\subsection{Semi-supervised Deep Generative Models for Reject Inference}\label{sec_ssdgm}
In reject inference, the data set $D = \{D_{accept},D_{reject}\}$ is composed of $n$ (labeled) accepted applications $D_{accept}=\{(\bm{x},y)_1,...,(\bm{x},y)_n\}$ and $m$ (unlabeled) rejected applications $D_{reject}=\{\mathbf{x}_{n+1}, ...,\mathbf{x}_{n+m}\}$, where $\mathbf{x} \in \mathbb{R}^{\ell_x}$ is the feature vector and $y_i \in \{0,1\}$ is the class label or the outcome of the loan, $y=0$ if the customer repaid the loan, otherwise $y=1$. Additionally, generative models assume that latent variable $\bm{z} \in \mathbb{R}^{\ell_z}$ governs the distribution of $\bm{x}$. 

The goal of the generative model is to obtain the joint distribution $p(\bm{x},y)$ of the data used for credit scoring and the outcome of the loan. However, this distribution is intractable since it requires integration over the whole latent space, i.e. $\int p(\bm{x},y,\bm{z}) d\bm{z}$. Further, the intractability of $p(\bm{x},y)$ translates into an intractable posterior distribution of $\bm{z}$ through the relationship
\begin{equation}
    p(\bm{z}|\bm{x},y) = \frac{p(\bm{x},y,\bm{z})}{\int p(\bm{x},y,\bm{z}) d\bm{z}}.
\end{equation}
Hence, we approximate the true posterior $p(\bm{z}|\bm{x},y)$ with the inference model $q(\bm{z}|\bm{x},y)$ and minimize the Kullback-Leibler (KL) divergence\footnote{The KL divergence is a measure of the proximity between two densities, e.g. $KL[q(\cdot)||p(\cdot)]$, and it is commonly measured in bits. It is non-negative and it is minimized when $q(\cdot)=p(\cdot)$.} $KL[q(\bm{z}|\bm{x},y)||p(\bm{z}|\bm{x},y)]$ to make the approximation as close as possible to the true density. 

The $KL[q(\bm{z}|\bm{x},y)||p(\bm{z}|\bm{x},y)]$ term, the objective  function $\mathcal{L}_{accept}$, and the density $p(x,y)$ are related by the following expression 
\begin{align}
    \log p(\bm{x},y)   &= \mathbb{E}_{q(\bm{z}|\bm{x},y)}[\log p(\bm{x},y)] \nonumber \\
    &= \mathbb{E}_{q(\bm{z}|\bm{x},y)}\Big[\log \frac{p(\bm{x},y,\bm{z})}{p(\bm{z}|\bm{x},y)} \frac{q(\bm{z}|\bm{x},y)}{q(\bm{z}|\bm{x},y)} \Big] \nonumber \\
    &= \mathbb{E}_{q(\bm{z}|\bm{x},y)}\Big[\log \frac{p(\bm{x},y,\bm{z})}{q(\bm{z}|\bm{x},y)} \Big] +\mathbb{E}_{q(\bm{z}|\bm{x},y)}\Big[\log \frac{q(\bm{z}|\bm{x},y)}{p(\bm{z}|\bm{x},y)} \Big] \nonumber \\
    &\vcentcolon= -\mathcal{L}_{accept}(\bm{x},y) + KL[q(\bm{z}|\bm{x},y)||p(\bm{z}|\bm{x},y)].
    \label{eq_sup_lb_dev}
\end{align}
Given that the $KL$ divergence in Equation \ref{eq_sup_lb_dev} is strictly positive, the term $-\mathcal{L}_{accept}(\bm{x},y)$ is a lower bound on $\log p(\bm{x},y)$, i.e. $\log p(\bm{x},y) \geq -\mathcal{L}_{accept}(\bm{x},y)$. Hence, since we cannot evaluate $p(\bm{z}|\bm{x},y)$, we maximize $\log p(\bm{x},y)$ by maximizing the negative lower bound.

Note that in Equation \ref{eq_sup_lb_dev} we assume that the outcome $y$ of the loan is known. However, this is not the case for the rejected applications $D_{reject}$. In this case, generative models treat $y$ as a latent variable and approximate the true posterior distribution $p(y|\bm{x})$ with the parametric function $q(y|\bm{x})$. Assuming the factorization $q(\bm{z},y|\bm{x})=q(y|\bm{x})q(\bm{z}|\bm{x},y)$ and a simple form for $q(y|\bm{x})$, we can take the explicit expectation over the class label $y$, i.e. we handle the uncertainty about the outcome of the loan by summing over the two possible outcomes that it might have taken. Mathematically,
\begin{align}
    \mathbb{E}_{q(\bm{z},y|\bm{x})}\Big[\log \frac{p(\bm{x},y,\bm{z})}{q(\bm{z},y|\bm{x})} \Big] &=
    \mathbb{E}_{q(y|\bm{x})}\mathbb{E}_{q(\bm{z}|\bm{x},y)}\Big[\log \frac{p(\bm{x},y,\bm{z})}{q(\bm{z},y|\bm{x})} \Big] \nonumber \\
    &= \mathbb{E}_{q(y|\bm{x})}[-\mathcal{L}_{accept}(\bm{x},y) - \log q(y|\bm{x})] \nonumber \\
    &= \sum_y q(y|\bm{x})[-\mathcal{L}_{accept}(\bm{x},y) - \log q(y|\bm{x})] \nonumber \\
    &\vcentcolon= -\mathcal{L}_{reject}(\bm{x}).
    \label{eq_obj_unsup}
\end{align}
Therefore, the objective function in semi-supervised deep generative models is the sum of the supervised lower bound for the accepted applications and the unsupervised lower bound for the rejected applications
\begin{equation}
    \mathcal{L} =  \mathcal{L}_{accept}(\bm{x},y)  + \mathcal{L}_{reject}(\bm{x}).
    \label{eq_objective}
\end{equation}
Furthermore, deep generative models parametrize the parameters of the density functions in Equation \ref{eq_sup_lb_dev} and \ref{eq_obj_unsup} by multilayer perceptron (MLP) networks. For example, if $\bm{z}|\bm{x},y$ is multivariate Gaussian distributed with diagonal covariance matrix, we use the notation
\begin{equation}
p(\bm{z}|\bm{x},y) \sim \mathcal{N}(\bm{z}|\bm{x},y; \ \bm{\mu} = f_{\bm{\theta}}(\bm{x},y), \bm{\sigma}^2\bm{I} = f_{\bm{\theta}}(\bm{x},y)),
\label{eq_notation_dgm}
\end{equation}
where $\bm{\mu} \in \mathbb{R}^{\ell_z}$ and $\bm{\sigma}^2 \in \mathbb{R}^{\ell_z}$, to specify that the parameters of the Gaussian distribution are parametrized by an MLP network denoted by $f(\bm{x},y)$ with input data $\bm{x},y$ and weights $\bm{\bm{\theta}}$\footnote{Deep generative models can also be developed with convolutional neural networks (CNNs). However, CNNs require structured data like videos, images, or time-series data. The data sets in this research are feature vectors with customer's characteristics at the application time. This kind of data does not have the grid-like structure required for training CNNs. For an application of CNNs in credit scoring the reader is referred to \cite{kvamme2018predicting}.}. Hence, the optimization of the objective function is with respect to the weights in the MLP. An alternative notation is to simply use the subscript $\bm{\theta}$ in the corresponding distribution, i.e. $p_{\bm{\theta}}(\bm{z}|\bm{x},y)$.
 
Finally, note that the EM algorithm used in \cite{anderson2013modified,feelders2000credit} cannot be used in this context since it requires to compute the expectation of $p(\bm{z}|\bm{x},y)$, which it is intractable. Other variational inference techniques, like mean-field or stochastic variational inference, determine different values of $\bm{\mu}_i$ and $\bm{\sigma}_i^2$ for each data point $\bm{x}_i$, which is computationally expensive. Similarly, traditional EM algorithms need to compute an expectation w.r.t the whole data set before updating the parameters. Therefore, deep generative models use complex functions of the data $\bm{x}$ (MLP networks) to estimate the best possible values for the latent variables $\bm{z}$. This allows replacing the optimization of point-specific parameters $\bm{\mu}_i$ and $\bm{\sigma}_i^2$, with a more efficient optimization of the MLP weights $\bm{\theta}$. The latter is denoted amortized inference \cite{zhang2018advances}.
 
\subsection{Model 1: Generative and inference process}\label{sec_model1}
\begin{figure}[t!]
    \centering
    \includegraphics[scale=1]{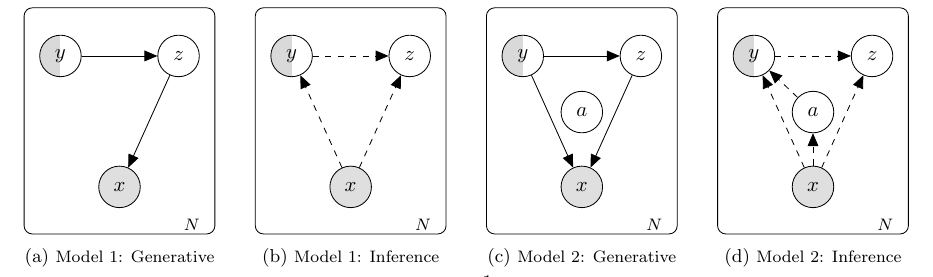}
    \caption{Plate notation for Model 1 and Model 2 where $\bm{x}$ is the observed feature vector, $y$ is the outcome of the loan and it is only observed for the accepted applications, and $\bm{z}$ and $\bm{a}$ are latent variables. The generative process is specified by solid lines, while the inference process is shown with dotted lines. Note that the MLP weights $\bm{\theta}$ and $\bm{\phi}$ lie outside the plates and we omit them to do not clutter the diagrams.}
\label{fig_platenotation}
\end{figure}
In this section we build upon the work done in \cite{kingma2014semi,zheng2016variational} to develop a new semi-supervised model with a Gaussian mixture parameterized with MLPs. The Gaussian mixture induces a flexible latent space that improves the approximation of the lower bound \citep{maaloe2016auxiliary,maaloe2015improving}. Hence, Model 1 assumes a generative process $p_{\bm{\theta}}(\bm{x},y,\bm{z})=p(y)p_{\bm{\theta}}(\bm{z}|y)p_{\bm{\theta}}(\bm{x}|\bm{z})$, where $\bm{x} \perp y  |  \bm{z}$, with the following probability density functions 
\begin{align}
    p(y) &\sim \text{Bernoulli}(y; \pi),  \nonumber \\
    p(\bm{z}|y) &\sim \mathcal{N}(\bm{z}|y=k; \ \bm{\mu}_{z_k} = f_{\bm{\theta}}(y), \bm{\sigma}_{z_k}^2\mathbf{I} = f_{\bm{\theta}}(y)) \ for \ k=0,1, \nonumber \\
    p(\bm{x}|\bm{z}) &\sim \mathcal{N}(\bm{x}|\bm{z}; \ \bm{\mu}_x = f_{\bm{\theta}}(\bm{z}), \bm{\sigma}_x^2\mathbf{I} = f_{\bm{\theta}}(\bm{z})). 
    \label{eq_gen_m1}
\end{align}
Here $\mathcal{N}$ denotes the Gaussian distributions and $f(\cdot)$ is a multilayer perceptron model with weights denoted by $\bm{\theta}$. Furthermore, we assume that the inference process is factorized as $q(\bm{z},y|\bm{x})=q(y|\bm{x})q(\bm{z}|\bm{x},y)$, with the following probability densities
\begin{align}
    q(y|\bm{x}) &\sim \text{Bernoulli}(y; \pi_{y|\bm{x}} =f_{\bm{\phi}}(\bm{x})),  \nonumber \\
    q(\bm{z}|\bm{x},y) &\sim \mathcal{N}(\bm{z}|\bm{x},y; \ \bm{\mu}_z = f_{\bm{\phi}}(\bm{x},y), \bm{\sigma}_z^2\mathbf{I} = f_{\bm{\phi}}(\bm{x},y)).
    \label{eq_varitaion_m1}
\end{align}
Again $\mathcal{N}$ is the Gaussian distribution and $f(\cdot)$ is a multilayer perceptron model with weights denoted by $\bm{\phi}$. Note that the marginal distribution $p(\bm{z})$ in the generative process is a GMM, i.e.
\begin{align*}
p(\bm{z}) =& \sum_y p(y)p(\bm{z}|y) \\
=& \pi \mathcal{N}(\bm{\mu}_{z_0}, \bm{\sigma}_{z_0}^2\mathbf{I}) + (1-\pi) \mathcal{N}(\bm{\mu}_{z_1}, \bm{\sigma}_{z_1}^2\mathbf{I}),
\end{align*}
where $(1-\pi)$ represents the prior for the default probability. The generative and inference processes are shown in Figure \ref{fig_platenotation}. 

In the following sections, we use $\bm{\theta}$ and $\bm{\phi}$ to distinguish the expectation and variance terms in the generative process from the ones in the inference process as well as to differentiate the MLP's weights in the generative process from the ones in the inference process. Further, we derive the lower bound for the supervised and unsupervised data under our novel approach for reject inference in credit scoring.

\subsubsection*{Labeled data: Deriving the objective function $\mathcal{L}_{accept}$}
\begin{figure}[t!]
    \centering
    \includegraphics[scale=0.7]{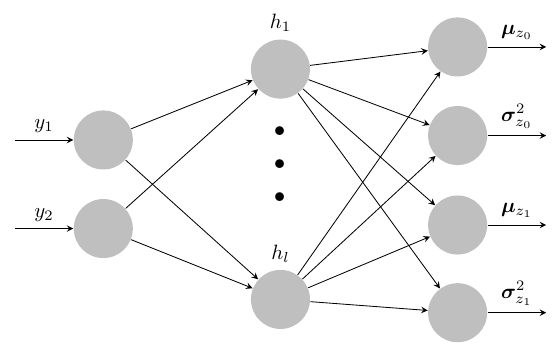}
    \caption{Gaussian mixture components parameterized by a multilayer perceptron model, where $y_{\cdot}$ is the one-hot-encoding for the input data ([$y_1 \ y_2$] = [0 1] and [$y_1 \ y_2$] = [1 0] are the one-hot-encoding for $y = 1$ and $y = 0$ respectively), $h_l$ is the \textit{l}'th neuron in the hidden layer, and $\bm{\mu}_{z_i}$ and $\bm{\sigma}_{z_i}$ are density moments for the \textit{i}'th component in the GMM. For the accepted applications, we backpropagate trough its corresponding component, while for the rejected applications we backpropagate through both components.}
    \label{fig_gmm}
\end{figure}

We use Equation \ref{eq_sup_lb_dev} and the factorization of the generative process in Equation \ref{eq_gen_m1} to derive the lower bound for the accepted data set $D_{accept}$. Hence, expanding the terms in the lower bound we obtain
\begin{align}
    \mathbb{E}_{q_{\bm{\phi}}(\bm{z}|\bm{x},y)}\Big[\log \frac{p_{\bm{\theta}}(\bm{x},y,\bm{z})}{q_{\bm{\phi}}(\bm{z}|\bm{x},y)} \Big] &=
    \mathbb{E}_{q_{\bm{\phi}}(\bm{z}|\bm{x},y)}[\log p(y) + \log p_{\bm{\theta}}(\bm{z}|y) + \log  p_{\bm{\theta}}(\bm{x}|\bm{z}) - \log  q_{\bm{\phi}}(\bm{z}|\bm{x},y)],
    \label{eq_sup_lb_m1}
\end{align}
and taking the expectations, see Section \ref{deriv_m1_sup} in the Appendix, we find the negative lower bound for a single (supervised) data point, which is 
\begin{align}
    - \mathcal{L}_{accept}(\{\bm{x},y\}_i;\bm{\theta},\bm{\phi})=\frac{1}{2}\Big[\sum_{j=1}^{\ell_z}(1+\log  \sigma_{\bm{\phi}_j}^2)- \sum_{j=1}^{\ell_z}\Big( \log \sigma_{\bm{\theta}_{j,y}}^2 + \frac{\sigma_{\bm{\phi}_j}^2}{\sigma_{\bm{\theta}_{j,y}}^2} +& \frac{(\mu_{\bm{\phi} _j}-\mu_{\bm{\theta}_{j,y}})^2}{\sigma_{\bm{\theta}_{j,y}}^2}\Big) \Big] + \log \pi_i \nonumber \\
    +&  \frac{1}{L} \sum_{l=1}^L \log \mathcal{N}(x_i| z_{i,l}). 
    \label{eq_sup_bound_m1}
\end{align}
Here $\ell_z$ is the dimension of $\bm{z}$, $\sigma_{\cdot_j}^2$ and $\mu_{\cdot_j}$ are the \textit{j}'th element of $\bm{\sigma}_{\cdot}^2$ and $\bm{\mu}_{\cdot}$ respectively, $\pi_i$ is the prior distribution over the class label $y_i$, and $L$ is the number of $\bm{z}_{i,l}$ samples drawn from $q_{\bm{\phi}}(\bm{z}|\bm{x},y)$. We use the \textit{reparametrization trick} $\bm{z}_{i,l}=\bm{\mu}_{i_{\bm{\phi}}} + \bm{\sigma}_{i_{\bm{\phi}}} \odot \bm{\epsilon}_l$, where $\bm{\epsilon}_l \sim \mathcal{N}(\bm{0},\mathbf{I})$ and $\odot$ denotes an element-wise multiplication, to backpropagate through $\bm{\sigma}_{\cdot}^2$ and $\bm{\mu}_{\cdot}$. Hence, the last term in Equation \ref{eq_sup_bound_m1} is $\mathcal{N}(x_i| z_{i,l}=\bm{\mu}_{i_{\bm{\phi}}} + \bm{\sigma}_{i_{\bm{\phi}}} \odot \bm{\epsilon}_l)$ and we use $q_{\bm{\phi}}(\bm{z}|\bm{x},y)$ to sample $\bm{\mu}_{i_{\bm{\phi}}}$ and $\bm{\sigma}_{i_{\bm{\phi}}}$. Note that since $y$ is known in this case, we only need to backpropagate through its corresponding Gaussian component in the MLP parameterizing the GMM. In other words, if $y_i=0$ the stochastic gradient optimization only updates all weights in $\bm{\mu}_{\bm{\theta}_y}$ and $\bm{\sigma}_{\bm{\theta}_y}^2$ for the first component in Figure \ref{fig_gmm}. This is specified by the subscript $y$ in Equation \ref{eq_sup_bound_m1}.

\subsubsection*{Unlabeled data: Deriving the objective function $\mathcal{L}_{reject}$}
In this case, we treat the unknown labels $y$ as latent variables and we approximate the true posterior distribution with $q(y|\bm{x})$. Given that $q(y|\bm{x}) \sim \text{Bernoulli}(\cdot)$ is a relatively easy distribution, we take the explicit expectation in the unsupervised lower bound. Following the steps in Equation \ref{eq_obj_unsup} together with the factorization in Equations \ref{eq_gen_m1} and \ref{eq_varitaion_m1}, we obtain
\begin{align}
    \mathbb{E}_{q_{\bm{\phi}}(\bm{z},y|\bm{x})}\Big[\log \frac{p_{\bm{\theta}}(\bm{x},y,\bm{z})}{q_{\bm{\phi}}(\bm{z},y|\bm{x})} \Big] &=
    \mathbb{E}_{q_{\bm{\phi}}(\bm{z},y|\bm{x})}[\log p(y) + \log p_{\bm{\theta}}(\bm{z}|y) + \log  p_{\bm{\theta}}(\bm{x}|\bm{z}) - \log q_{\bm{\phi}}(y|\bm{x}) \nonumber \\ 
    & \hspace{7.1cm} - \log q_{\bm{\phi}}(\bm{z}|\bm{x},y)] \nonumber \\
    &= \mathbb{E}_{q_{\bm{\phi}}(y|\bm{x})}[-\mathcal{L}_{accept}(\bm{x};\bm{\theta},\bm{\phi}) - \log q_{\bm{\phi}}(y|\bm{x})] \nonumber \\
    &= \sum_y q_{\bm{\phi}}(y|\bm{x})[-\mathcal{L}_{accept}(\bm{x};\bm{\theta},\bm{\phi}) - \log q_{\bm{\phi}}(y|\bm{x})],
    \label{eq_objm1_unsup}
\end{align}
which is, by definition, the unsupervised negative lower bound $- \mathcal{L}_{reject}(\bm{x};\bm{\theta},\bm{\phi})$. Furthermore, taking the expectations, see Section \ref{deriv_m1_unsup} in the Appendix, we can obtain the negative lower bound for a single data point, which is
\begin{align}
    - \mathcal{L}_{reject}(\bm{x}_i;\bm{\theta},\bm{\phi})=\frac{1}{2} \sum_{y=0}^1 \pi_{y|\bm{x}_i} \Big[\sum_{j=1}^{\ell_z}(1 +& \log \sigma_{\bm{\phi}_j}^2)- \sum_{j=1}^{\ell_z}\Big( \log \sigma_{\bm{\theta}_{j,y}}^2 + \frac{\sigma_{\bm{\phi}_j}^2}{\sigma_{\bm{\theta}_{j,y}}^2} + \frac{(\mu_{\bm{\phi} _j}-\mu_{\bm{\theta}_{j,y}})^2}{\sigma_{\bm{\theta}_{j,y}}^2}\Big) \Big] \nonumber \\
    &+\sum_{y=0}^1 \pi_{y|\bm{x}_i} \log \frac{\pi}{\pi_{y|\bm{x}_i}} +  \frac{1}{L} \sum_{l=1}^L \log \mathcal{N}(\bm{x}_i| z_{i,l}), 
\end{align}
where $\pi_{y|\bm{x}}$ is the \textit{y}'th element of the posterior probability over the class labels $\bm{\pi}_{y|\bm{x}}=[\pi_{y=0|\bm{x}} \ (1-\pi_{y=0|\bm{x}})]$. The rest of the parameters have the same interpretation as in the supervised negative lower bound. Note that in this case we take the expectation over the latent variable $y$ by enumerating the two possible values ($y=0$ and $y=1$) of the posterior parameter $\bm{\pi}_{y|\bm{x}}$, which also implies that we need to backpropagate through the two components, one at a time, in $\bm{\sigma}_{\bm{\theta}_y}^2$ and $\bm{\mu}_{\bm{\theta}_y}$, see Figure \ref{fig_gmm}.

We train Model 1 alternating the objective function 
\begin{equation}
    \mathcal{L} =  \sum_i^n \mathcal{L}_{accept}\big((\bm{x},y)_i;\bm{\theta},\bm{\phi}\big) - \alpha \cdot \log \mathbb{E}_{\hat{p}(\bm{x},y)}[q_{\bm{\phi}}(y_i|\bm{x}_i)] + \sum_j^{n+m} \mathcal{L}_{reject}(\bm{x}_j;\bm{\theta},\bm{\phi}),
    \label{eq_objective_m1}
\end{equation}
where $\mathbb{E}_{\hat{p}(\bm{x},y)}$ is the empirical distribution.

Note that we introduce the term $\log \mathbb{E}_{\hat{p}(\bm{x},y)}[q_{\bm{\phi}}(y_i|\bm{x}_i)]$, which is actually the classifier in Model 1, into the supervised lower bound to take advantage of the accepted applications and train the best possible classifier. The term $\alpha = \beta \cdot \frac{m+n}{n}$ controls the importance of the classification in the supervised loss function, where $m$ and $n$ are the number of rejected and accepted observations respectively, and $\beta$ is just a scaling factor.

\subsubsection{Reject Inference in Credit Scoring with Model 1}\label{sec_understaningM1}
Model 1 does not just learn the distribution $p(\bm{x}|\bm{z})$ of the customers' data used in credit scoring, but it also learns a latent representation $p(\bm{z}|\bm{x},y)$ of it. This latent representation reflects an intrinsic structure or the semantics of the customers' data. Additionally, Model 1 approximates the posterior class label distribution $q(y|\bm{x})$, which we use to estimate the default probability for new applications. This probability is given by the mutually exclusive outcomes in the posterior parameter $\bm{\pi}_{y|\bm{x}}$, which is parametrized by an MLP with softmax activation function in the output layer.

The most important characteristic of Model 1 for reject inference in credit scoring is that the unknown creditworthiness is evaluated by considering the two possible states $y=1$ and $y=0$ that the loan might have taken in case that the credit had been granted (Equation \ref{eq_objm1_unsup}). This means that this method clearly differs from all extrapolation approaches for reject inference. Further, it is not as restrictive as the expectation-maximization algorithm since it relies on the approximation of the posterior distributions. 

It can be shown that Equation \ref{eq_objective_m1} includes the term $KL[q_{\bm{\phi}}(\bm{z}|\bm{x},y)||p_{\bm{\theta}}(\bm{z}|y)]$. Then, the optimization of the objective function forces $q_{\bm{\phi}}(\bm{z}|\bm{x},y)$ to be as close as possible to $p_{\bm{\theta}}(\bm{z}|y)$, which we have modeled as a mixture of Gaussian distributions. The first motivation for this is that the data for the accepted and rejected applications are generated by two different process, just as in \cite{feelders2000credit}. Second, this mixture model generates a flexible latent space, which helps to improve the approximation of the inference process in Model 1.

Finally, the objective function in Equation \ref{eq_objective_m1} includes the MLP weights $\bm{\theta}$ for the densities $p(\bm{z}|y)$ and $p(\bm{x}|\bm{z})$, and $\bm{\phi}$ for the densities $q(y|\bm{x})$ and $q(\bm{z}|\bm{x},y)$. These are all the weights in Model 1 and are present in both the supervised and unsupervised loss. Hence, the stochastic gradient optimization updates these weights jointly and estimates the different parameters $\bm{\mu}$, $\bm{\sigma}^2$, and $\bm{\pi}$ in Equation \ref{eq_gen_m1} and \ref{eq_varitaion_m1}. In practice, when a labeled (accepted) observation is presented to the algorithm, the loss function in the backpropagation algorithm is   $\mathcal{L}_{accept}\big((\bm{x},y)_i;\bm{\theta},\bm{\phi}\big)$. Similarly, when handling unlabeled (rejected) observations the loss function is $\mathcal{L}_{reject}(\bm{x}_j;\bm{\theta},\bm{\phi})$. In any case, all the MLP weights $\bm{\theta}$ and $\bm{\phi}$ are updated at each iteration since the same MLP handles both accepted and rejected applications.

\subsection{Model 2: Generative and inference processes}\label{sec_model2}
Inspired by the work by \cite{maaloe2015improving,maaloe2016auxiliary}, we develop an extension of Model 1 introducing auxiliary variables. Auxiliary variables improve the variational approximation and introduce a layer of latent variables to the model's classifier. Hence, our proposed Model 2 combines a Gaussian mixture with auxiliary variables in a semi-supervised framework for the first time in the literature.

Specifically, we assume the generative process $p(\bm{x},y,\bm{z},\bm{a})=p(\bm{a})p(y)p(\bm{z}|y)p(\bm{x}|\bm{z},y)$ with the following distributions
\begin{align}
    p(y) &\sim \text{Bernoulli}(y;\pi),   \nonumber \\
    p(\bm{a}) &\sim \mathcal{N}(\bm{a}; \ \bm{0}, \bm{1}), \nonumber \\ 
    p(\bm{z}|y) &\sim \mathcal{N}(z|y=k; \ \bm{\mu}_{z_k} = f_{\bm{\theta}}(y), \bm{\sigma}_{z_k}^2\mathbf{I} = f_{\bm{\theta}}(y)) \ for \ k=0,1, \nonumber \\ 
    p(\bm{x}|\bm{z},y) &\sim \mathcal{N}(\bm{x}|\bm{z},y; \ \bm{\mu}_x = f_{\bm{\theta}}(\bm{z},y), \bm{\sigma}_x^2\mathbf{I} = f_{\bm{\theta}}(\bm{z},y)).
    \label{eq_generative_m2}
\end{align}
Here $\mathcal{N}$ is the Gaussian distribution and $f(\cdot)$ is a multilayer perceptron model with weights denoted by $\bm{\theta}$. The inference process factorizes as $q(\bm{z},\bm{a},y|x)=q(\bm{a}|\bm{x})q(y|\bm{x},\bm{a})q(\bm{z}|\bm{x},y)$. The distributions for this process are 
\begin{align}
    q(\bm{a}|\bm{x}) &\sim \mathcal{N}(\bm{a}|\bm{x}; \ \bm{\mu}_a = f_{\bm{\phi}}(\bm{x}), \bm{\sigma}_a^2\mathbf{I} = f_{\bm{\phi}}(\bm{x})), \nonumber \\
    q(y|\bm{x},\bm{a}) &\sim \text{Bernoulli}(y|\bm{x},\bm{a}; \ \bm{\pi}_{y|\bm{x},\bm{a}} = f_{\bm{\phi}}(\bm{x},\bm{a})), \nonumber \\
    q(\bm{z}|\bm{x},y) &\sim \mathcal{N}(\bm{z}|\bm{x},y; \ \bm{\mu}_z = f_{\bm{\phi}}(\bm{x},y), \bm{\sigma}_z^2\mathbf{I} = f_{\bm{\phi}}(\bm{x},y)).
    \label{eq_inf_m2}
\end{align}
Again $\mathcal{N}$ is the Gaussian distribution and $f(\cdot)$ is a multilayer perceptron model with weights denoted by $\bm{\phi}$.

\subsubsection*{Labeled data: Deriving the objective function $\mathcal{L}_{accept}$}
Following the steps in Section \ref{sec_ssdgm}, it is straightforward to show that the supervised negative lower bound  is 
\begin{align}
    - \mathcal{L}(\bm{x},y;\bm{\theta},\bm{\phi})_{accept}&=\mathbb{E}_{q_{\bm{\phi}}(\bm{z},\bm{a}|\bm{x},y)}\Big[ \log \frac{p_{\bm{\theta}}(\bm{x},y,\bm{z},\bm{a})}{q_{\bm{\phi}}(\bm{z},\bm{a}|\bm{x},y)}\Big] \nonumber \\
    &= \mathbb{E}_{q_{\bm{\phi}}(\bm{z},\bm{a}|\bm{x},y)} [\log p(\bm{a}) + \log p(y) + \log p_{\bm{\theta}}(\bm{z}|y) + \log  p_{\bm{\theta}}(\bm{x}|\bm{z},y) \nonumber \\ 
    &- \log q_{\bm{\phi}}(\bm{a}|\bm{x}) - \log q_{\bm{\phi}}(\bm{z}|\bm{x},y)].
\end{align}
Using Equations \ref{eq_generative_m2} and \ref{eq_inf_m2} and taking the corresponding expectations, see Section \ref{deriv_m2_sup} in the Appendix, we obtain the lower bound for the \textit{i}'th data point, as follows\footnote{We clutter the notation by adding the subscript $\bm{a}$ and $\bm{z}$ in the distribution parameters. This helps to differentiate the parameters of the density $q_{\bm{\phi}}(\bm{a}|\bm{x})$ from the ones in $q_{\bm{\phi}}(\bm{z}|\bm{x},y)$.}
\begin{align}
    - \mathcal{L}_{accept}((\bm{x},y)_i;\bm{\theta},\bm{\phi}) =& \frac{1}{2}\Big[\sum_{j=1}^{\ell_z}(1+\log  \sigma_{\bm{\phi}_{\bm{z}_j}}^2) - \sum_{j=1}^{\ell_z}\Big( \log \sigma_{\bm{\theta}_{j,y}}^2 + \frac{\sigma_{\bm{\phi}_{\bm{z}_j}}^2}{\sigma_{\bm{\theta}_{j,y}}^2} + \frac{(\mu_{\bm{\phi}_{\bm{z}_j}}-\mu_{\bm{\theta}_{j,y}})^2}{\sigma_{\bm{\theta}_{j,y}}^2}\Big) \Big] + \log \pi_i \nonumber \\
    +& \frac{1}{2} \sum_{c=1}^{\ell_a}(\sigma_{\bm{\phi}_{\bm{a}_c}}^2 + \mu_{\bm{\phi}_{\bm{a}_c}}^2 - (1+\log  \sigma_{\bm{\phi}_{\bm{a}_c}}^2))
    + \frac{1}{L_z} \sum_{l=1}^{L_z} \log \mathcal{N}(x_i|z_{i,l},y).
    \label{eq_sup_bound_m2}
\end{align}
Here $\ell_z$ and $\ell_a$ are the dimensions of $\bm{z}$ and $\bm{a}$ respectively, $\sigma_{\cdot_j}^2$ and $\mu_{\cdot_j}$ are the \textit{j}'th element of $\bm{\sigma}_{\cdot}^2$ and $\bm{\mu}_{\cdot}$ respectively, and they refer to the variance or expectation of either $\bm{z}$ or $\bm{a}$, $\pi_i$ is the prior distribution over the class label $y_i$, and $L_z$ is the number of $\bm{z}_{i,l}$ samples drawn from $q_{\bm{\phi}}(\bm{z}|\bm{x},y)$. Note that $y$ is known in this case, hence we only backpropagate through its corresponding Gaussian component, just as in Model 1. This is specified by the subscript $y$ in Equation \ref{eq_sup_bound_m2}.

\subsubsection*{Unlabeled data: Deriving the objective function $\mathcal{L}_{reject}$}
Using the factorization in Equation \ref{eq_generative_m2} and \ref{eq_inf_m2}, the unsupervised negative lower bound in Model 2 has the form
\begin{align}
    -\mathcal{L}_{reject}(\bm{x};\bm{\theta},\bm{\phi}) &= \mathbb{E}_{q_{\bm{\phi}}(\bm{z},\bm{a},y|\bm{x})}\Big[ \log  \frac{p_{\bm{\theta}}(\bm{x},y,\bm{z},\bm{a})}{q_{\bm{\phi}}(\bm{z},\bm{a},y|\bm{x})}\Big] \nonumber \\
    &= \mathbb{E}_{q_{\bm{\phi}}(\bm{z},\bm{a},y|\bm{x})} [\log p(\bm{a}) + \log p(y) + \log p_{\bm{\theta}}(\bm{z}|y) + \log  p_{\bm{\theta}}(\bm{x}|\bm{z},y) \nonumber \\ 
    &- \log q_{\bm{\phi}}(\bm{a}|\bm{x}) - \log q_{\bm{\phi}}(\bm{z}|\bm{x},y) - \log q_{\bm{\phi}}(y|\bm{x},\bm{a})].
    \label{eq_m2_unsup}
\end{align}
For the \textit{i}'th observation, Equation \ref{eq_m2_unsup} takes the following form, see Section \ref{deriv_m2_unsup} in the Appendix, 
\begin{align}
    - \mathcal{L}_{reject}(\bm{x}_i;\bm{\theta},\bm{\phi}) =& \frac{1}{2}\frac{1}{L_a}\frac{1}{L_z}\sum_{l_a=1}^{L_a} \sum_{y=0}^1 \pi_{y|\bm{x}_i,\bm{a}_{i,l_a}} \bigg[\sum_{j=1}^{\ell_z}(1+\log \sigma_{\bm{\phi}_{\bm{z}_j}}^2) - \sum_{j=1}^{\ell_z}\bigg( \log \sigma_{\bm{\theta}_{j,y}}^2 + \frac{\sigma_{\bm{\phi}_{\bm{z}_j}}^2}{\sigma_{\bm{\theta}_{j,y}}^2} \nonumber \\
    +& \frac{(\mu_{\bm{\phi}_{\bm{z}_j}}-\mu_{\bm{\theta}_{j,y}})^2}{\sigma_{\bm{\theta}_{j,y}}^2}\bigg) 
    +\frac{1}{L_z} \sum_{l_z=1}^{L_z} \log \mathcal{N}(\bm{x}_i| \bm{z}_{i,l_z},y_{l_a})\bigg] + \frac{1}{2} \sum_{c=1}^{\ell_a} \Big(\sigma_{\bm{\phi}_{\bm{a}_c}}^2 + \mu_{\bm{\phi}_{\bm{a}_c}}^2 \nonumber \\
    -& (1+\log \sigma_{\bm{\phi}_{\bm{a}_c}}^2)\Big) + \frac{1}{L_a}\sum_{l_a=1}^{L_a} \sum_{y=0}^1 \pi_{y|\bm{x}_i,\bm{a}_{i,l_a}}(-\log q(y|\bm{x}_i,\bm{a}_{i,l_a})) + \log \pi_i.
\end{align}
Here all parameters are just as in $- \mathcal{L}_{accept}(\bm{x},y;\bm{\theta},\bm{\phi})$. It is important to note that the posterior probability over the class labels $\bm{\pi}_{y|\bm{x},\bm{a}}=[\pi_{y=0|\bm{x},\bm{a}} \ (1-\pi_{y=0|\bm{x},\bm{a}})]$  depends on the sampled auxiliary variables. We denote this dependency explicitly using the subscript $\bm{a}$.

Finally, just as we did in Model 1, we include the term $\log q_{\bm{\phi}}(y|\bm{x},\bm{a})$ in the unsupervised objective function to take advantage of the accepted applications. Therefore, the final objective function for Model 2 is \begin{equation}
    \mathcal{L} =  \sum_i^m \mathcal{L}_{accept}\big((\bm{x},y)_i;\bm{\theta},\bm{\phi}\big) - \alpha \cdot \log \mathbb{E}_{\hat{p}(\bm{x},y,\bm{a})}[q_{\bm{\phi}}(y_i|\bm{x}_i,\bm{a}_i)] + \sum_j^n \mathcal{L}_{reject}(\bm{x}_j;\bm{\theta},\bm{\phi}).
    \label{eq_objective_m2}
\end{equation}

\subsubsection{Reject Inference in Credit Scoring with Model 2}\label{sec_understaningM2}
Model 2 has almost the same characteristics as Model 1, but there are two new items. First, Model 2 approximates two layers of latent representations $q(\bm{a}|\bm{x})$ and $q(\bm{z}|\bm{x},y)$. The posterior distribution $q(\bm{a}|\bm{x})$, together with the customers' data $\bm{x}$, is used to estimate the default probability (Equation \ref{eq_inf_m2}). By doing so, Model 2 has a relatively more expressive estimation of creditworthiness. The presumption is that the latent representation $\bm{a}$ captures the intrinsic structure of the data and that it therefore provides relevant features for enhancing the performance of the classifier $q(y|\bm{x},\bm{a})$. Finally, note that $q(\bm{a}|\bm{x})$ is assumed to be multivariate Gaussian distributed, hence we use the reparametrization trick (see Section \ref{sec_model1}) to sample from this distribution, i.e. $\bm{a} =\bm{\mu}_{_a} + \bm{\sigma}_{a} \odot \bm{\epsilon}$ where $\bm{\mu}_{a}$ and $\bm{\sigma}_{a}$ are the outputs in the MLP for the density $q(\bm{a}|\bm{x})$.

The second difference from Model 1 is that the data generating process $p(\bm{x}|\bm{z},y)$ is conditioned on the latent variable $\bm{z}$ and class label $y$. This is simply done to achieve better training stability. See Section \ref{sec_implementation} for more details about model training.   

\section{Experiments and Results}\label{sec_results}
\begin{figure}[t!]
    \centering
    \includegraphics[scale=0.5]{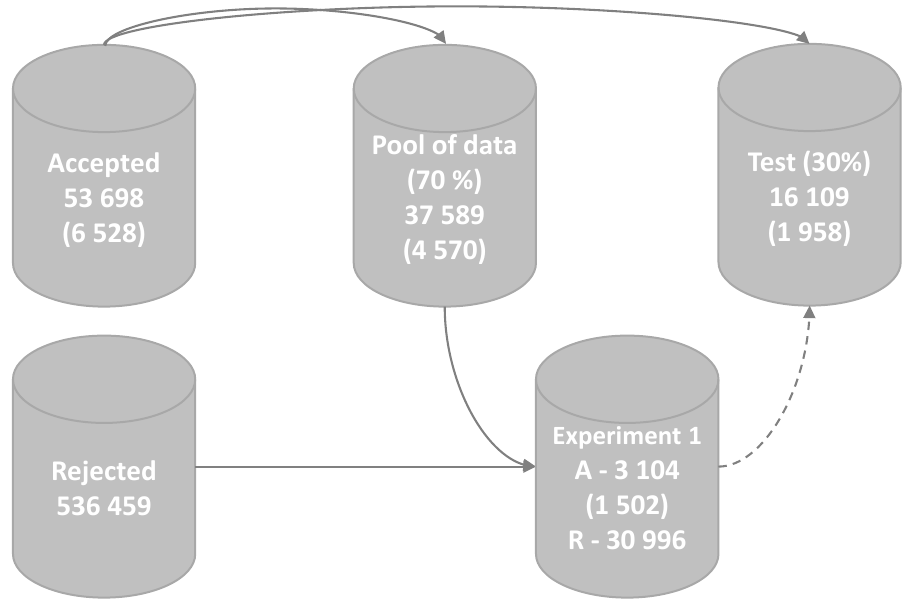}
    \caption{Data partition used in the experiments in Table \ref{tbl_results} for the Lending Club data set. Numbers in parentheses are the number of defaulted observations, and numbers in parenthesis in percentage are the proportion of accepted applications. The experiments with the Santander data set and in Table \ref{tbl_results4} follows the same logic, but in the last sampling ('Experiment 1' box) we sample the number of accepted and rejected applications as needed.}
    \label{fig_data_partition}
\end{figure}
The goal with the experiments is twofold. First, we compare the performance of our proposed models with a range of techniques representing the state-of-the-art in reject inference for credit scoring, including three classical reject inference techniques (reclassification, fuzzy parceling and augmentation \cite{hsia1978credit}) and three semi-supervised machine learning approaches (self-learning \cite{rosenberg2005semi} MLP, self-learning SVM, and semi-supervised SVM \cite{gieseke2012sparse}) under a realistic scenario preserving the original acceptance rates in two real data sets. Second, to have a better understanding of the behaviour of reject inference models for credit scoring, we test the model performance in different scenarios varying the number of accepted and rejected observations. In both cases, we include two supervised machine learning models (multilayer perceptron (MLP) \cite{rumelhart1985learning} and support vector machine (SVM) \cite{cortes1995support}) to measure the marginal gain of reject inference\footnote{The code for Model 1 and Model 2 can be found at \href{https://github.com/rogelioamancisidor/reject_inference}{https://github.com/rogelioamancisidor/reject\_inference}}.

\subsection{Data description}
We use two real data sets containing both rejected and accepted applications. The first data set is public\footnote{The data can be obtain directly at the Lending Club's website, however they require the user to login. We obtain a complete version of the available data at the website \href{https://github.com/nateGeorge/preprocess_lending_club_data}{$\text{https://github.com/nateGeorge/preprocess\_lending\_club\_data}$}, which is updated quarterly.} and consists of personal loan applications through Lending Club, which is the world's largest peer-to-peer lending company. We replicate the data sample used in \cite{li2017reject}, which includes applications from January 2009 until September 2012 with 36-months maturity. However, we do not split the data set in yearly sub samples, since we want to keep as many observations from the minority class ($y=1$) as possible. Hence, the data set that we use in our experiments has 53 698 accepted applications, including 6 528 defaults, and 536 459 rejected applications\footnote{The number of accepted and rejected applications are not exactly the same as in \cite{li2017reject}, but the variable statistics are very similar and the default trend is the same. See Table \ref{tbl_lc_statistics} for more information.}. That is, the acceptance ratio is 9.10\% and default rate is 12.16\%. For more details about the Lending Club data, see Table \ref{tbl_lc_statistics} in the Appendix.

The second data set is provided by Santander Consumer Bank Nordics and consists of credit card applications arriving trough their internet website. The applications were received during the period January 2011 until December 2016. During this period Santander accepted 126 520 applications and only 14 993 customers ended up as defaults. The number of rejected applications during this period is 232 898. Hence, the acceptance ratio is 35.20\% and default rate 11.85\%.

In addition to these two data sets, we have two small samples after September 2012 and December 2016 for Lending Club and Santander Bank respectively, which are used to produce well-calibrated estimates of class probabilities using the beta calibration approach \cite{kull2017beta}. These samples are not part of the experimental design explained in Section \ref{sec_experiments}.

\begin{table}[t!]
\centering 
\begin{adjustbox}{width=\textwidth}
\begin{tabular}{|l|l|}
\hline
\multicolumn{2}{|c|}{Lending Club and Santander Credit Cards}\\
\hline 
MLP Network &  \multicolumn{1}{c|}{Number of hidden layers and dimensions}             \\
\hline
$q(\bm{z}|\bm{x},y)$& [10 10]*, [10 20], [10 30], [10 50], [100 70]***, [10 20 10], [10 30 10], [10 40 10]**, [10 50 10], [60 90 60]****      \\
$p(\bm{x}|\cdot)$   & [10 10]*, [10 20], [10 30], [10 50], [70 100]***, [10 20 10], [10 30 10], [10 40 10]**, [10 50 10], [60 90 60]****	   \\
$p(\bm{z}|y)$	    & $\text{[10]}^{*,**,***,****}$	                                             \\
$q(\bm{a}|\bm{x})$  & 	[50], [10 10], [10 20], [10 30], [10 40]**, [10 50], [20 40], [20 50], [30 50], [30 60], [40 60]****	 \\ 
$q(y|\cdot)$        & [50], [60], [70]*,[80]***, [100]****, [120], [130]**   \\
\hline
Parameter/hyperparameter & \multicolumn{1}{c|}{Value} \\
\hline
$\bm{z}$ dimension  & 30, $\text{50}^{*,**,****}$, 100***  \\
$\bm{a}$ dimension  & 30, $\text{50}^{**,****}$  \\
$\beta$	& 0.008**, 0.01, 0.025, 0.14, 1.1*, 3****, 8*** \\
\hline
\end{tabular}
\end{adjustbox}
\caption{Grid for hyperparameter optimization for Model 1 and 2 and for both data sets. The numbers within brackets specify the number of neurons in each hidden layers, i.e. $[10 \ 10]$ means two hidden layers with 10 neurons each. Finally, the superscript * and ** shows the final architecture for Model 1 and Model 2 respectively for the Lending Club data set used in Table \ref{tbl_results}. Similarly, *** and **** shows the final architecture for Model 1 and Model 2 respectively for the Santander Credit Cards data set used in Table \ref{tbl_results}.}
\label{tbl_gridsearch_m1m2_sa}
\end{table}

\subsection{Experimental Design}\label{sec_experiments}
We conduct two different set of experiments. In the first experimental setup, we keep the original acceptance ratio, but we do not use more than 34 100 observations in total \footnote{This is done to allow a fair comparison to S3VM, which does not scale to larger datasets due to memory requirements. For the 34 100 observations, S3VM requires 123GB of memory to estimate the kernel matrix.}. To construct this data set, we first split the original data in 70\%-30\% for training and testing respectively. Then, we down sample the majority class ($y=0$) in the training set until it equals the number of observations for the minority class ($y=1$). To achieve the correct acceptance ratio, this requires a random selection of both class labels. Note that the test data set is left as it is, i.e. it preserves the original default rate. Finally, we randomly select the number of reject applications in a way that these, together with the balanced training sample, do not exceed 34 100 observations, see Figure \ref{fig_data_partition}. 

In the second set of experiments\footnote{S3VM is not included in this section since it takes around 356 hours to evaluate each scenario in this section and in total we evaluate 12 different scenarios. In addition, it has the memory restrictions already mentioned. Similarly, the iterative procedure in the self-learning SVM is not feasible in this section.}, we analyze the effect of varying the number of accepted (rejected) applications, while keeping the same number of rejected (accepted) applications. We follow the same approach as in the the first experiments, splitting the data set into a training and test data set, down sampling the training set, and randomly selecting the number of reject applications. 

For the Lending Club data set, we use all variables in Table \ref{tbl_lc_statistics} to train all models, while for the Santander data we use a forward selection approach to select the explanatory variables that are included in the reclassification, fuzzy parceling and augmentation methods\footnote{These three methods are based on the logistic regression. Hence, the forward selection approach prevents the logistic regression from overfitting and avoids numerical problems on its optimization.}. For the other models we use all variables in Table \ref{tbl_santander_statistics}. Finally, we do hyperparameter tuning using grid search with 10-cross validation for the MLP, SVM, S3VM, Model 1, and Model 2. The best architecture for the MLP and SVM is used as the base model in the self-training approaches for MLP and SVM. The details of the grid search are given in Table \ref{tbl_gridsearch}.

\npthousandsep{}
\npdecimalsign{.}
\nprounddigits{4}
\begin{table}[t!]
\centering 
\begin{adjustbox}{width=\textwidth}
\begin{tabular}{|l|n{1}{4}n{1}{4}n{1}{4}n{1}{4}n{1}{4}|n{1}{4}n{1}{4}n{1}{4}n{1}{4}n{1}{4}|n{1}{4}|n{1}{4}|}
\hline
 & \multicolumn{5}{c|}{Lending Club (LC) } & \multicolumn{5}{c|}{Santander Credit Cards (SCC) } &  \multicolumn{2}{c|}{Runtime} \\
\hline 
& {AUC} & {GINI} & {H-measure} & {Recall}  & {Precision}& {AUC} & {GINI} & {H-measure} & {Recall}  & {Precision} & \multicolumn{1}{c|}{LC} & \multicolumn{1}{c|}{SCC} \\
\hline
MLP                     & 0.6273437	& 0.2546876 & 0.05352209 & 0.4453524 & 0.173809049 &  0.7091459&0.4182919&0.13257518&0.790929301911961&0.177222426109141 & \multicolumn{1}{c|}{00:01.28} & \multicolumn{1}{c|}{00:04.53} \\
SVM                     & 0.6283748	& 0.2567493	& 0.05431159 & 0.463227783 &0.178332678 &0.7388437&0.4776869&0.1689229&0.799733214762116&0.189529684054504 & \multicolumn{1}{c|}{00:06.59} & \multicolumn{1}{c|}{00:14.42} \\
\hline
Reclassification        & 0.5783648&0.1567293&0.02273701&0.490551583248212&0.14925181632906& 0.6414885&0.282977&0.06249591&0.998888394842152&0.118703723970146  & \multicolumn{1}{c|}{00:05.04} & \multicolumn{1}{c|}{00:01.15} \\
Fuzzy Parceling         & 0.619804&0.2559605&0.05402885&0.459805924412666&0.177153499338293&0.6790781&0.3581565&0.09569927&0.867630057803468&0.154121610672991 & \multicolumn{1}{c|}{00:03.82} & \multicolumn{1}{c|}{00:08.45} \\
Augmentation            & 0.6218834&0.2557669&0.0540965&0.458120531154239&0.177694937740762& 0.6761321&0.3522645&0.09230714&0.873521565140062&0.152446758065431 & \multicolumn{1}{c|}{00:13.07} & \multicolumn{1}{c|}{00:15.25} \\
\hline
Self-lerning MLP        & 0.586849&0.17369783&0.032602819&0.450357507660878&0.157031027737485&  0.6725641&0.3451283&0.08768281&0.850244553134726&0.151859300613936 & \multicolumn{1}{c|}{00:18.80} & \multicolumn{1}{c|}{00:20.53} \\
Self-lerning SVM        & 0.6205615&0.2551228&0.05346849&{\npboldmath}0.495709908069459&0.173126365532489& 0.7266105&0.4532207&0.1529293&0.849444197421076&0.172548002678795  & \multicolumn{1}{c|}{03:25.89} & \multicolumn{1}{c|}{05:08.36} \\
S3VM                    & 0.6201002&0.2402004&0.04814717&0&\multicolumn{1}{c|}{NA}&0.6519966&0.3039933&0.0732743&{\npboldmath}1&0.118505638107282  & \multicolumn{1}{c|}{09:17.00} & \multicolumn{1}{c|}{06:20.12} \\
\hline
Model 1                 & 0.629388895176  &  0.258777790353  &  0.0554185026503  & 0.45398365679 & 0.178807071233 & 0.739419092381&0.478838184762&0.167805819644& 0.832592263226 & 0.184774028024 & \multicolumn{1}{c|}{10:48.19} & \multicolumn{1}{c|}{04:12.16} \\
Model 2        & {\npboldmath}0.6362851401477376         &   {\npboldmath} 0.275464084381 & {\npboldmath} 0.0632190280788 &   0.468845760978  &  {\npboldmath} 0.182496471062 & {\npboldmath}  0.743148668624  &  {\npboldmath} 0.485097337248  &  {\npboldmath} 0.176410955592 & 0.628166740772  &  {\npboldmath} 0.23027528913 & \multicolumn{1}{c|}{12:24.06} & \multicolumn{1}{c|}{05:54.33} \\
\hline
\end{tabular}
\end{adjustbox}
\caption{Model performance keeping the original acceptance ratios, i.e. 9.10\% for Lending Club (LC) and 35.20\% for Santander Credit Cards (SCC). The training data set is balanced by down sampling the majority class, and the threshold used to calculate recall and precision is based on the empirical default rate in the test data set. The last two columns show the runtime for one cross-validation and the format is given in mm:ss.cs, where mm, ss, and cs stands for minutes, seconds and centiseconds respectively.}
\label{tbl_results}
\end{table}

\begin{figure}[t!]
    \centering
    \includegraphics[scale=0.59]{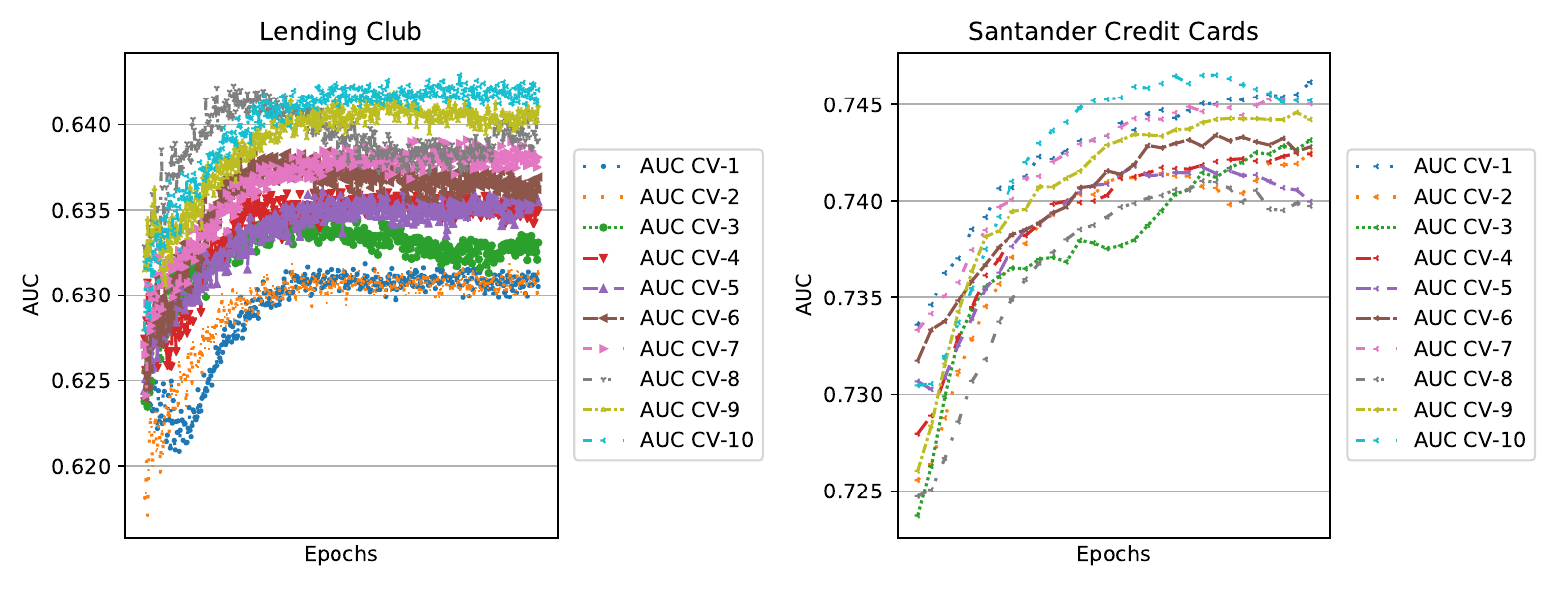}
    \caption{The left panel shows the AUC performance for the Lending Club data set in the 10 cross-validations (CV), while the right panel shows the performance for the Santander Bank data set. Both diagrams correspond to Model 2.}
    \label{fig:my_label}
\end{figure}

\subsection{Model Implementation and Training}\label{sec_implementation}
Model 1 and Model 2 are implemented in Theano \cite{2016arXiv160502688short}. We use softplus activation functions in all hidden layers and linear activation functions in all output layers estimating $\bm{\mu}$ and $\bm{\sigma}^2$. For the output layer in the classifiers $q_{\bm{\phi}}(y|\cdot)$ we use softmax activation functions. Further, we use the Adam optimizer \cite{kingma2014adam} with learning rate equal to 1e-4 and 5e-5 for training of Model 1 and Model 2 respectively. The rest of parameters in the Adam optimizer are the default values suggested in the original paper. We use $L=1$ and $L_a=1$ for both Model 1 and 2 in all experiments. Finally, both data sets are standardized before training and testing, and the class label $y$ is one-hot-encoded. The model architectures used in the experiments in Table \ref{tbl_results} are shown in Table \ref{tbl_gridsearch_m1m2_sa}.

It is important to mention that deep generative models are, in general, difficult to train \cite{liu2018analyzing,takahashi2018student}. The training of Model 1 and Model 2 in some cases become unstable, especially for the experiments where we vary the number of accepted and rejected applications. Moreover, it is sensitive to the initial weights. Hence, we use a Variational Autoencoder \cite{kingma2013auto} to pretrain the weights in $q_{\bm{\phi}}(\bm{z}|\bm{x},y)$ and $p_{\bm{\theta}}(\bm{x}|\bm{z})$ for Model 1. Similarly, we prewarm all weights $\bm{\theta}$ and $\bm{\phi}$ in Model 2. In both cases, we initialized the MLP weights as suggested in \cite{glorot2010understanding}. We also achieve more stable training in Model 2 by conditioning $p_{\bm{\theta}}(\bm{x}|\bm{z},y)$ on the class label $y$.  

\subsection{Benchmark Reject Inference}\label{sec_orig_ratios}
\begin{table}[t!]
\centering 
\begin{adjustbox}{width=\textwidth}
\begin{tabular}{|l|n{1}{4}n{1}{4}n{1}{4}n{1}{4}n{1}{4}n{1}{4}|n{1}{4}n{1}{4}n{1}{4}n{1}{4}n{1}{4}n{1}{4}|}
\hline
 \multicolumn{13}{|c|}{Lending Club} \\
\hline  
 & \multicolumn{6}{c|}{Accepted applications} & \multicolumn{6}{c|} {Rejected applications} \\
\hline  
\multirow{2}{*}{No. observations} & \multicolumn{1}{c}{200} & \multicolumn{1}{c}{600} & \multicolumn{1}{c}{1 200} & \multicolumn{1}{c}{2 000}  & \multicolumn{1}{c}{6 000} & \multicolumn{1}{c|}{All} & \multicolumn{1}{c}{30 997} & \multicolumn{1}{c}{100 000} & \multicolumn{1}{c}{200 000} & \multicolumn{1}{c}{300 000}  & \multicolumn{1}{c}{400 000} & \multicolumn{1}{c|}{All} \\
 & \multicolumn{1}{c}{(0.04\%)} & \multicolumn{1}{c}{(0.11\%)}& \multicolumn{1}{c}{(0.22\%)}& \multicolumn{1}{c}{(0.37\%)}& \multicolumn{1}{c}{(1.11\%)}& \multicolumn{1}{c|}{(1.67\%)}& \multicolumn{1}{c}{(0.64\%)} & \multicolumn{1}{c}{(0.20\%)}& \multicolumn{1}{c}{(0.10\%)}& \multicolumn{1}{c}{(0.07\%)}& \multicolumn{1}{c}{(0.05\%)}& \multicolumn{1}{c|}{(0.04\%)} \\
\hline
MLP                     & 0.6001927&  0.62363353& 0.62366077& 0.63043247& 0.62986954 &0.63074811  & 0.60370998 &  0.60370998& 0.60370998& 0.60370998& 0.60370998 &0.60370998   \\
SVM                     & 0.60388527& 0.62674405& 0.62527006 &0.63203706& 0.63016335& 0.63091588  & 0.605363962& 0.605363962& 0.605363962 &0.605363962& 0.605363962& 0.605363962 \\
\hline
Reclassification        & 0.57861003& 0.57845679& 0.58117156& 0.58533913 &0.58055825& 0.58157524 & 0.56160402 &0.57852433& 0.57830356& 0.55738218& 0.5692937 & 0.57794348 \\
Fuzzy Parceling         & 0.60174793& 0.62399504& 0.62317408 &0.62948137& 0.62965029 &0.63022516  & 0.60412932 &0.60256955& 0.60184787& 0.60310012& 0.60730376& 0.60062518 \\  
Augmentation            & 0.60166053 &0.6215873 & 0.62065128& 0.63006822 &0.6294754&  0.63035666  & 0.6023336  &0.60281574& 0.60100891& 0.59673167& 0.59526177& 0.59785979  \\
\hline
Self-lerning MLP        & 0.5823714 & 0.572756 & 0.5733693 & 0.567486 & 0.5858446 & 0.5630668     & 0.5639674	& 0.5484652	& 0.5705547 & 0.5714778 &	0.5758168 &	0.5702585  \\
\hline
Model 2                 &    {\npboldmath}    0.6174538380466909    &  {\npboldmath}0.6269250429165094     & {\npboldmath} 0.6310491049388773      &  {\npboldmath}  0.6344088961752481   &  {\npboldmath} 0.6380687393863558     &  {\npboldmath} 0.6403639516332523    &    {\npboldmath} 0.6112450412521052    & {\npboldmath} 0.6075210587544769     &   {\npboldmath} 0.6091223530488475    &   {\npboldmath} 0.6106655598066267  &   {\npboldmath} 0.612082119795525  &   {\npboldmath} 0.6174538380466909      \\
\hline
Runtime & & & & & & & & & & & & \\
\hline
Self-learning MLP &\multicolumn{1}{c}{00:20:36}&\multicolumn{1}{c}{00:26:14}&\multicolumn{1}{c}{00:29:31}&\multicolumn{1}{c}{00:29:23}&\multicolumn{1}{c}{00:31.39}&\multicolumn{1}{c|}{00:35:11}&\multicolumn{1}{c}{00:02.10}&\multicolumn{1}{c}{00:05:02}&\multicolumn{1}{c}{00:09:50}&\multicolumn{1}{c}{00:15:01}&\multicolumn{1}{c}{00:18:02}&\multicolumn{1}{c|}{00:23:36}\\
Model 2 &\multicolumn{1}{c}{02:39:02}&\multicolumn{1}{c}{02:41:75}&\multicolumn{1}{c}{02:55:19}&\multicolumn{1}{c}{03:24:13}&\multicolumn{1}{c}{03:42:17}&\multicolumn{1}{c|}{04:03:10}&\multicolumn{1}{c}{00:14:18}&\multicolumn{1}{c}{00:38:07}&\multicolumn{1}{c}{01:09:02}&\multicolumn{1}{c}{01:39:48}&\multicolumn{1}{c}{02:00:54}&\multicolumn{1}{c|}{02:39:02}\\
\hline
\end{tabular}
\end{adjustbox}
\caption{Left panel: Model performance, measured with AUC, as a function of accepted applications. In all six experiments to the left, we use all 536 459 rejected applications. Right panel: Model performance, measured with AUC, as a function of rejected applications. In all six experiments to the right, we use only 200 accepted applications. Numbers in parenthesis are the acceptance ration for each experiment. The last two rows show the runtime for one cross-validation and the format is in hh:mm:ss, where hh, mm, and ss stands for hours, minutes, and seconds respectively. We do not include the runtime for the first five models because the difference with respect to the runtimes in Table \ref{tbl_results} is negligible.}
\label{tbl_results4}
\end{table}

Table \ref{tbl_results} compares the performance of Model 1 and Model 2 with other models when using the original acceptance ratio in the data sets. It can be seen that both Model 1 and Model 2 perform better than all supervised and semi-supervised models in terms of AUC, GINI, H measure and precision. Our results support previous findings that the reclassification, fuzzy parcelling and augmentation methods do not improve model performance. The reclassification approach is consistently the worst model. Further, the self-training approaches do not improve the performance of the base models MLP and SVM. Finally, S3VM has significantly worse performance than the base models for the Santander Credit Cards data set.  

We use the Platt scaling method \cite{platt1999probabilistic} to get (pseudo) default probabilities from SVM and S3VM. It is interesting to see that we could not estimate the recall and precision for S3VM in the Lending Club data because the estimated default probabilities are concentrated around the average, with practically no dispersion, see Table \ref{tbl_results2}. S3VM estimates default probabilities for all applications below the default rate in the Lending Club data set, and above the default rate in the Santander data set. 

Model 2 performs better than Model 1 in terms of all measures except for recall. Remember that the main difference between these models is the classifier in Model 2, which uses a latent representation of the customers' data. Our results are hence in correspondence with previous studies showing the predictive power embedded in the latent transformations. It is further interesting to note that our proposed models for reject inference not only perform better, but also estimate higher variability in the predicted default probabilities, as shown in Table \ref{tbl_results2}. This result supports previous findings that the default probability is underestimated if reject inference is ignored. Unfortunately, given the nature of the data sets in this research we are not able to draw any conclusion about the economic impact of this interesting detail. 

It is worth mentioning that Model 2 is the algorithm that takes longer time to converge for the Lending Club data set, while for the Santander Credit Cards data set is S3VM. In any case, the runtime for both Model 2 and S3VM, in the experiments in Table \ref{tbl_results}, is moderate.

In Table \ref{tbl_results4}, we analyze the impact of the number of accepted and rejected applications on model performance using Model 2 and the Lending Club data set. In the right panel, we can observe that the general trend is that the more rejected applications we add to Model 2, the higher model performance. In the left panel, we can see that the more accepted data we have available, the better model performance for the supervised models and the less difference compared to Model 2. Note that Model 2 achieves the highest average AUC of 0.6404 in the \textit{All} scenario, which includes 545 599 observations. This is 16 times more data compared to what self-training SVM and S3VM handled. 

The runtime for Model 2 in the experiments that use all rejected applications has increased significantly compared to Table \ref{tbl_results}. In the scenario where we use all accepted and rejected applications, 545 599 observations in total, Model 2 takes about 4 hours to converge. Note that this model has 16 080 learnable parameters, which are significantly more than the 502 parameters in the MLP. Generally, training deep learning architectures is computationally intensive and the computational complexity increases linearly with the number of parameters (including MLP architectures). However, training can be accelerated by distributing training in parallel across multiple GPUs.

\section{Conclusion}\label{sec_conclusion}
In this research we develop two novel deep generative models for reject inference in credit scoring. Our models use the posterior distribution of the outcome of the loan to infer the unknown creditworthiness of the rejected applications. This is done by exact enumeration of the two possible outcomes of the loan, which is an advantage compared to reject inference methods based on extrapolation. To the best of our knowledge, this is the first research that develops novel methods for reject inference in credit scoring coupling Gaussian mixtures and auxiliary variables in a semi-supervised framework with generative models.

The experiments show that our proposed models achieve higher model performance compared to many of the classical and machine learning approaches for reject inference in credit scoring, and the models' performance increases as we add more data for model training. Further, the efficient stochastic gradient optimization technique used in deep generative models scales to large data sets, which is an advantage over supervised and semi-supervised support vector machines. Note that even though the focus of this research is on credit scoring, our proposed models generalize to other research domains, e.g. image classification. 

The higher model performance of our proposed methodology is further enhanced by adding latent representations of the customers' data to the classifier. This data representation captures the intrinsic structure of the data providing relevant information for classification. Since our proposed approach for reject inference in credit scoring offers flexible modeling possibilities, we hope that this research spurs future work on reject inference in credit scoring using deep generative model focusing on improving the training stability and classification power.  

\section*{Acknowledgements}
The authors would like to thank Santander Consumer Bank for financial support and the real data set used in this research. This work was also supported by the Research Council of Norway [grant number 260205] and SkatteFUNN [grant number 276428].

\clearpage
\section{Appendix} \label{sec_appendix}
\renewcommand{\thefigure}{A\arabic{figure}}
\setcounter{figure}{0}
\renewcommand{\theequation}{B\arabic{equation}}
\setcounter{equation}{0}
\renewcommand{\thetable}{A\arabic{table}}
\setcounter{table}{0}
\addcontentsline{toc}{section}{Appendices}
\renewcommand{\thesubsection}{\Alph{subsection}}

\subsection{Tables and Figures}

\npdecimalsign{.}
\nprounddigits{2}
\npthousandsep{\hspace{0.3em}}
To replicate the data set presented in \cite{li2017reject}, we excluded all observations with missing values in any of the variables in Table \ref{tbl_lc_statistics}. Further, the allowed variable range, which we choose based on \cite{li2017reject}, is determined by the minimum and maximum values as shown in the table. The summary statistics in our data sample is not exactly the same as in \cite{li2017reject}, but the default trend  is the same (the  default rate in 2009 is 12.59\%, 2010 is 9.61\%, 2011 is 10.32\% and in 2012 is 13.76\%).

\begin{table}[h]
\centering 
\caption{Lending Club Descriptive Statistics}
\begin{adjustbox}{width=0.8\textwidth}
\begin{tabular}{lln{5}{2}n{5}{2}n{4}{2}n{4}{2}n{5}{2}n{5}{2}n{5}{2}}
\hline
&{Variable}&{Mean}&{Std}&{Min}&{1 Quantile}&{Median}&{3 Quantile}&{Max} \\
\hline
Accepts&Debt to income&14.5133&7.1864&0&9.06&14.44&19.82&34.99 \\
&Loan amount&10610.3379&6738.6089&1000&5706.25&9600&14000&35000 \\
&Fico score&711.4917&35.0587&662&682&707&732&847.5 \\
&State d1&0.4276&0.4946&0&0&0&1&1 \\
&State d2&0.4265&0.4945&0&0&0&1&1 \\
&State d3&0.096&0.2946&0&0&0&0&1 \\
&Employment length&3.9747&3.1817&0&1&3&6&10 \\
\\
Rejects&Debt to income&24.2874&31.1397&0&7.9&18.19&31.18&419.33\\
&Loan amount&13330.7383&10361.5146&1000&5000&10000&20000&35000\\
&Fico score&638.1453&74.0992&385&595&651&690&850\\
&State d1&0.4679&0.4982&0&0&0&1&1\\
&State d2&0.3746&0.484&0&0&0&1&1\\
&State d3&0.101&0.3018&0&0&0&0&1\\
&Employment length&8.4008&3.1551&0&10&10&10&10\\
\hline
\end{tabular}
\end{adjustbox}
\label{tbl_lc_statistics}
\end{table}

The second data set which we use in this research is provided by Santander Consumer Bank. The details that we can provide about this data set are limited by its proprietary nature. The descriptive statistics are shown in Table \ref{tbl_santander_statistics}.

\begin{table}[h!]
\centering 
\caption{Santander Credit Cards Descriptive Statistics}
\label{tbl_santander_statistics}
\begin{adjustbox}{width=\textwidth}
\begin{tabular}{lln{6}{2}n{8}{2}n{10}{2}n{6}{2}n{6}{2}n{7}{2}n{10}{2}}
\hline
&{Variable}&{Mean}&{Std}&{Min}&{1 Quantile}&{Median}&{3 Quantile}&{Max} \\
\hline
Accepts&Var1&86475.8359&107975.2188&0&29852&69162&108898&10570323\\
&Var2&152205.1094&1778838.75&0&0&0&4376&393676928\\
&Var3&38.9506&13.3785&19&28&37&48&92\\
&Var4&976647.6875&16125692&-2&-2&-2&1250000&2701061888\\
&Var5&903518.75&3228558.75&-2&-2&-2&1430000&985694976\\
&Var6&807869.625&13848935&0&0&0&1075000&2667096064\\
&Var7&95622.1641&14090133&-2664925952&-2&-2&79000&984075008\\
&Var8&9.4613&23.8183&-2&-2&-2&4.63&100\\
&Var9&-0.4381&1.8595&-2&-2&-2&1&82\\
&Var10&-0.9148&1.1444&-2&-2&-2&0&4\\
&Var11&-1.993&0.1468&-2&-2&-2&-2&3\\
&Var12&-0.628&2.0582&-2&-2&-2&1&164\\
&Var13&-0.3353&2.0921&-2&-2&-2&1&164\\
&Var14&-1.9756&0.3216&-2&-2&-2&-2&26\\
&Var15&-0.4745&1.7345&-2&-2&-2&1&52\\
&Var16&-1.1549&0.9961&-2&-2&-2&0&1\\
&Var17&0.1623&0.5264&0&0&0&0&19\\
&Var18&0.9533&2.2475&0&0&0&1&67\\
&Var19&1.1157&2.4183&0&0&0&1&72\\
&Var20&1.5715&3.2728&0&0&0&2&97\\
&Var21&357123.8438&372109.8125&0&170103.1367&295917.4375&443333.9531&34850852\\
&Var22&8.2874&8.5329&0&3.9704&6.9129&10.2898&760.9356\\
&Var23&37156.3789&250887.75&-12873071&-14218.188&23241.04&79463.8242&33829372\\
&Var24&16168.6982&432254.875&-40114780&0&0&0&50003248\\
&Var25&9037.9854&60101.1719&-2641216&-4085&5520&19799.25&6169685\\
&Var26&0.3549&42.0393&0&0.1956&0.2345&0.2609&14940.1992\\
\hline
\end{tabular}
\end{adjustbox}
\end{table}

\begin{table}[ht!]
\centering 
\begin{adjustbox}{width=\textwidth}
\begin{tabular}{lln{6}{2}n{8}{2}n{10}{2}n{6}{2}n{6}{2}n{7}{2}n{10}{2}}
&&&&&&&&{Table 2 Continued}\\
\hline
&{Variable}&{Mean}&{Std}&{Min}&{1 Quantile}&{Median}&{3 Quantile}&{Max} \\
\hline
&Var27&0.4719&0.4993&0&0&0&1&1\\
&Var28&46.0447&75.6998&-29&-2&12&65&754\\
&Var29&6.7073&34.722&-2&-2&-2&-2&412\\
&Var30&6.7073&34.722&-2&-2&-2&-2&412\\
&Var31&1.0763&0.9656&0&0.5292&0.8989&1.3557&43.7516\\
&Var32&0.9829&1.0237&0&0.4715&0.8171&1.2214&101.9501\\
&Var33&0.9763&1.0088&0&0.4691&0.8118&1.216&99.1294\\
&Var34&0.5557&1.1805&0&0&0&1&73\\
&Var35&0.4901&0.4998&0&0&0&1&1\\
&Var36&0&0.0056&0&0&0&0&1\\
&Var37&0.5771&0.4944&0&0&1&1&1\\
&Var38&0.0661&0.2484&0&0&0&0&1\\
&Var39&0.208&0.4057&0&0&0&0&1\\
&Var40&0.093&0.2905&0&0&0&0&1\\
&Var41&0.0558&0.2295&0&0&0&0&1\\
&Var42&0.0119&0.1083&0&0&0&0&1\\
&Var43&0.3722&0.4836&0&0&0&1&1\\
&Var44&0.5281&0.4993&0&0&1&1&1\\
&Var45&0.0879&0.2832&0&0&0&0&1\\
&Var46&0.0005&0.0225&0&0&0&0&1\\
&Var47&0.6511&0.4764&0&0&1&1&1\\
&Var48&0.2575&0.4373&0&0&0&1&1\\
&Var49&0.056&0.2297&0&0&0&0&1\\
&Var50&0&0.004&0&0&0&0&1\\
&Var51&0.0349&0.1835&0&0&0&0&1\\
&Var52&0.7458&0.4353&0&0&1&1&1\\
&Var53&0.2542&0.4353&0&0&0&1&1\\
&Var54&0.0807&0.2723&0&0&0&0&1\\
&Var55&0.156&0.3629&0&0&0&0&1\\
&Var56&0.3852&0.4868&0&0&0&1&1\\
&Var57&0.2972&0.4569&0&0&0&1&1\\
&Var58&0.081&0.2727&0&0&0&0&1\\
\\
Rejects&Var1&57198.23046875&68931.4609375&0&12800&43182.5&80412&3635832\\
&Var2&33128.0078125&568171.5&0&0&0&0&208626176\\
&Var3&34.6028518676757&12.1608209609985&1&25&32&42&95\\
&Var4&507337.6875&11648304&-2&-2&-2&105937.5&2701061888\\
&Var5&434133.625&1137152.75&-2&-2&-2&0&72376000\\
&Var6&432619.65625&10198556&0&0&0&0&2303705088\\
&Var7&1499.87854003906&10159168&-2299855104&-2&-2&-2&72376000\\
&Var8&3.44621610641479&16.6995277404785&-2&-2&-2&0&100\\
&Var9&-1.16019892692565&1.50700807571411&-2&-2&-2&0&82\\
&Var10&-1.39079773426055&1.01886999607086&-2&-2&-2&0&4\\
&Var11&-1.86711776256561&0.949085235595703&-2&-2&-2&-2&36\\
&Var12&-1.24286592006683&1.67429280281066&-2&-2&-2&-2&105\\
&Var13&-1.06166219711303&1.77250158786773&-2&-2&-2&1&105\\
&Var14&-1.78726303577423&1.19774961471557&-2&-2&-2&-2&38\\
&Var15&-1.13481438159942&1.52476227283477&-2&-2&-2&1&43\\
&Var16&-1.51610147953033&0.869858920574188&-2&-2&-2&-2&1\\
&Var17&0.259388238191604&0.744392931461334&0&0&0&0&87\\
&Var18&3.27850818634033&6.06369495391845&0&0&1&4&166\\
&Var19&3.53789639472961&6.29928588867187&0&0&1&4&172\\
&Var20&4.61559581756591&7.90024662017822&0&0&2&5&176\\
&Var21&250519.140625&242146.78125&0&112918.59375&212571.75&337357.2890625&13897584\\
&Var22&5.80109930038452&5.55350875854492&0&2.63998389244079&4.93809556961059&7.84278106689453&308.835205078125\\
&Var23&23313.23828125&179360.1875&-31086966&-15761.4868164062&16862.4541015625&61574.015625&11590733\\
&Var24&2551.0390625&171498.015625&-30644804&0&0&0&16552538\\
&Var25&5758.37646484375&43678.19140625&-6499649&-3843&3537&14794&1851795\\
&Var26&0.295368313789367&31.142967224121&0&0.164311833679676&0.227324269711971&0.261938519775867&14940.19921875\\
&Var27&0.253707617521286&0.434827923774719&0&0&0&1&1\\
&Var28&32.2353363037109&65.4182968139648&-43&-2&-2&43&804\\
&Var29&6.67259502410888&32.3206748962402&-2&-2&-2&-2&377\\
&Var30&6.67259502410888&32.3206748962402&-2&-2&-2&-2&377\\
&Var31&0.773236513137817&0.697626590728759&0&0.349189788103103&0.668800950050354&1.04523980617523&36.9947433471679\\
&Var32&0.691182136535644&0.666885077953338&0&0.312046609818935&0.587580293416976&0.931432753801345&38.1634178161621\\
\hline
\end{tabular}
\end{adjustbox}
\end{table}

\begin{table}[ht]
\centering 
\begin{adjustbox}{width=\textwidth}
\begin{tabular}{lln{6}{2}n{8}{2}n{10}{2}n{6}{2}n{6}{2}n{7}{2}n{10}{2}}
&&&&&&&&{Table 2 Continued}\\
\hline
&{Variable}&{Mean}&{Std}&{Min}&{1 Quantile}&{Median}&{3 Quantile}&{Max} \\
\hline
&Var33&0.68791115283966&0.662265956401824&0&0.309911891818046&0.586197167634963&0.926732718944549&38.902774810791\\
&Var34&0.364408463239669&1.0669800043106&0&0&0&0&97\\
&Var35&0.273441582918167&0.445401906967163&0&0&0&1&1\\
&Var36&0.00017174900858663&0.013103081844747&0&0&0&0&1\\
&Var37&0.511412739753723&0.499769240617752&0&0&1&1&1\\
&Var38&0.0713660046458244&0.257168173789978&0&0&0&0&1\\
&Var39&0.230023443698883&0.42117565870285&0&0&0&0&1\\
&Var40&0.137064293026924&0.344136953353881&0&0&0&0&1\\
&Var41&0.0499617867171764&0.218047127127647&0&0&0&0&1\\
&Var42&0.00425937538966536&0.0650676935911178&0&0&0&0&1\\
&Var43&0.196665495634078&0.3976631462574&0&0&0&0&1\\
&Var44&0.746292352676391&0.434827923774719&0&0&1&1&1\\
&Var45&0.0527827627956867&0.223794668912887&0&0&0&0&1\\
&Var46&0.00044654740486294&0.0211225971579551&0&0&0&0&1\\
&Var47&0.742831647396087&0.437273830175399&0&0&1&1&1\\
&Var48&0.162139654159545&0.36806645989418&0&0&0&0&1\\
&Var49&0.0583302564918994&0.233779534697532&0&0&0&0&1\\
&Var50&0.058744988363469&0.00293042045086622&0&0&0&0&1\\
&Var51&0.0362433344125747&0.187034517526626&0&0&0&0&1\\
&Var52&0.547990977764129&0.497717022895813&0&0&1&1&1\\
&Var53&0.45200902223587&0.497717022895813&0&0&0&1&1\\
&Var54&0.0916796177625656&0.288628965616226&0&0&0&0&1\\
&Var55&0.159503296017646&0.366134017705917&0&0&0&0&1\\
&Var56&0.376263439655303&0.484214067459106&0&0&0&1&1\\
&Var57&0.280655056238174&0.44887426495552&0&0&0&1&1\\
&Var58&0.0889015793800354&0.28446826338768&0&0&0&0&1\\
\hline
\end{tabular}
\end{adjustbox}
\end{table}

\begin{table}[t!]
\centering 
\caption{Grid for hyperparameter optimization for Lending Club: The total number of model configurations are 132, 160 and 240 for MLP, SVM, and S3VM respectively. For the Santander data set the number of model configurations evaluated are 204, 160, and 240 for MLP, SVM, and S3VM respectively.}
\begin{adjustbox}{width=\textwidth}
\begin{tabular}{llllll}
\multicolumn{6}{c}{Lending Club} \\
\hline
\multicolumn{2}{c}{MLP} & \multicolumn{2}{c}{SVM} & \multicolumn{2}{c}{S3VM} \\
\hline 
Layers&1&C&5, 10, 13, 14, 15, 17, 19, 21, 23, 25&C&1, 5, 10, 13, 15, 17\\
Neurons&3, 15, 20, 25, 30, 35, 40, 45, 50, 55, 60&Gamma&2, 1.5, 1, 0.5, 0.1, 0.01, 0.001, auto&Gamma&2.5, 2, 1.5, 1, 0.5\\
Activation &logistic, tanh, relu&Kernel&rbf, linear&Kernel&rbf, linear\\
Learning rate&constant, adaptive&&&LamU&0.5, 1, 1.5, 2\\
Solver&sgd, adam&&&&\\
\multicolumn{6}{c}{Santander Credit Cards} \\
\hline
Layers&1&C&5, 10, 13, 14, 15, 17, 19, 21, 23, 25&C&1, 5, 10, 13, 15, 17\\
Neurons&50, 60, 65, 70, 75, 80, 85, 90, 95, 100, 105, 110, 115, 120, 130, 140, 150&Gamma&2, 1.5, 1, 0.5, 0.1, 0.01, 0.001, auto&Gamma&2.5, 2, 1.5, 1, 0.5\\
Activation &logistic, tanh, relu&Kernel&rbf, linear&Kernel&rbf, linear\\
Learning rate&constant, adaptive&&&LamU&0.5, 1, 1.5, 2\\
Solver&sgd, adam&&&&\\
\hline
\end{tabular}
\end{adjustbox}
\label{tbl_gridsearch}
\end{table}

\clearpage
\npthousandsep{}
\npdecimalsign{.}
\nprounddigits{4}
\begin{table}[ht!]
\footnotesize
\centering 
\caption{Empirical moment statistic for the default probability. }
\begin{adjustbox}{width=\textwidth}
\begin{tabular}{|l|n{1}{4}n{1}{4}n{1}{4}n{1}{4}|n{1}{4}n{1}{4}n{1}{4}n{1}{4}|}
\hline
 & \multicolumn{4}{c|}{Lending Club} & \multicolumn{4}{c|}{Santander Credit Cards} \\
\hline 
& \multicolumn{1}{c}{Average} & \multicolumn{1}{c}{Std.} & \multicolumn{1}{c}{Kurtosis} & \multicolumn{1}{c|}{Skewness}  & \multicolumn{1}{c}{Average} & \multicolumn{1}{c}{Std.} & \multicolumn{1}{c}{Kurtosis} & \multicolumn{1}{c|}{Skewness}  \\
\hline
MLP                     & 0.110083&0.0096088&-0.102672&0.0968617&0.118024&0.0146406&-0.0884836&0.0562915 \\
SVM                     & 0.101221&0.0130132&-0.150509&0.042008&0.120248&0.019894&-0.101564&0.0517189 \\
\hline
Reclassification        & 0.106638&0.00832874&-0.0635009&-0.286105& 0.120018&0.00107289&6.17302&-0.82066\\
Fuzzy Parceling         & 0.100334&0.0131776&-0.138887&0.0813118&0.119788&0.00411983&0.640579&-0.606143 \\  
Augmentation            & 0.09954&0.0130696&-0.148654&0.0880737&0.119799&0.00396129&0.628466&-0.615098\\
\hline
Self-learning MLP        & 0.105494&0.0116349&-0.0471382&0.0769663&0.127565&0.00577276&0.228217&-0.517921\\
Self-learning SVM        & 0.101445&0.0130059&-0.149382&0.038407& 0.125704&0.0146891&-0.119854&-0.0741083 \\
S3VM                    & 0.12032&\text{1.39e-6}&-0.117311&-0.129713&0.11995715&\text{7.08e-7}&0.740688&0.86868  \\
\hline
Model 1                 &  0.09846792855548125  &  0.04079455291155461  &  -0.565011181532 & 0.336769193798 & 0.11900005998867315  &  0.03668840520914141  &  -1.14592021122 &  -0.245477303994 \\
Model 2                 &   0.09992208570174972 & 0.042434932074115174 & -0.53662305745& 0.38188359808  &0.09250065237897293  &  0.03401597734050494  &  0.818187499143 &  0.780178298612 \\
\hline
\end{tabular}
\end{adjustbox}
\label{tbl_results2}
\end{table}

\begin{figure}[b!]
    \centering
    \includegraphics[angle=90,scale=0.41]{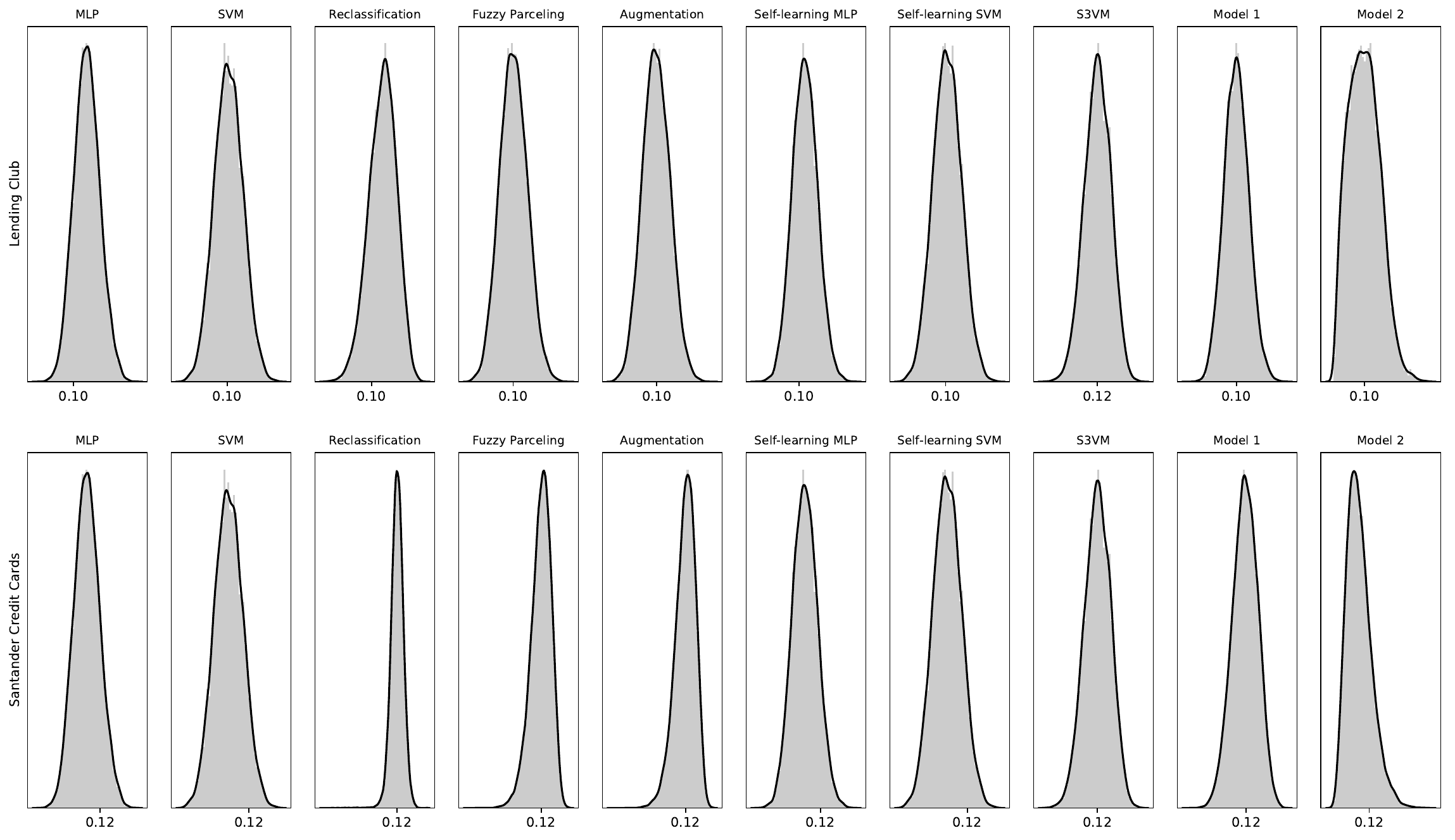}
    \caption{Empirical distribution of the default probability for the original acceptance ratio as explained in Section \ref{sec_experiments}.}
    \label{fig_distribution}
\end{figure}

\begin{figure}[h!]
    \centering
    \includegraphics[scale=0.7]{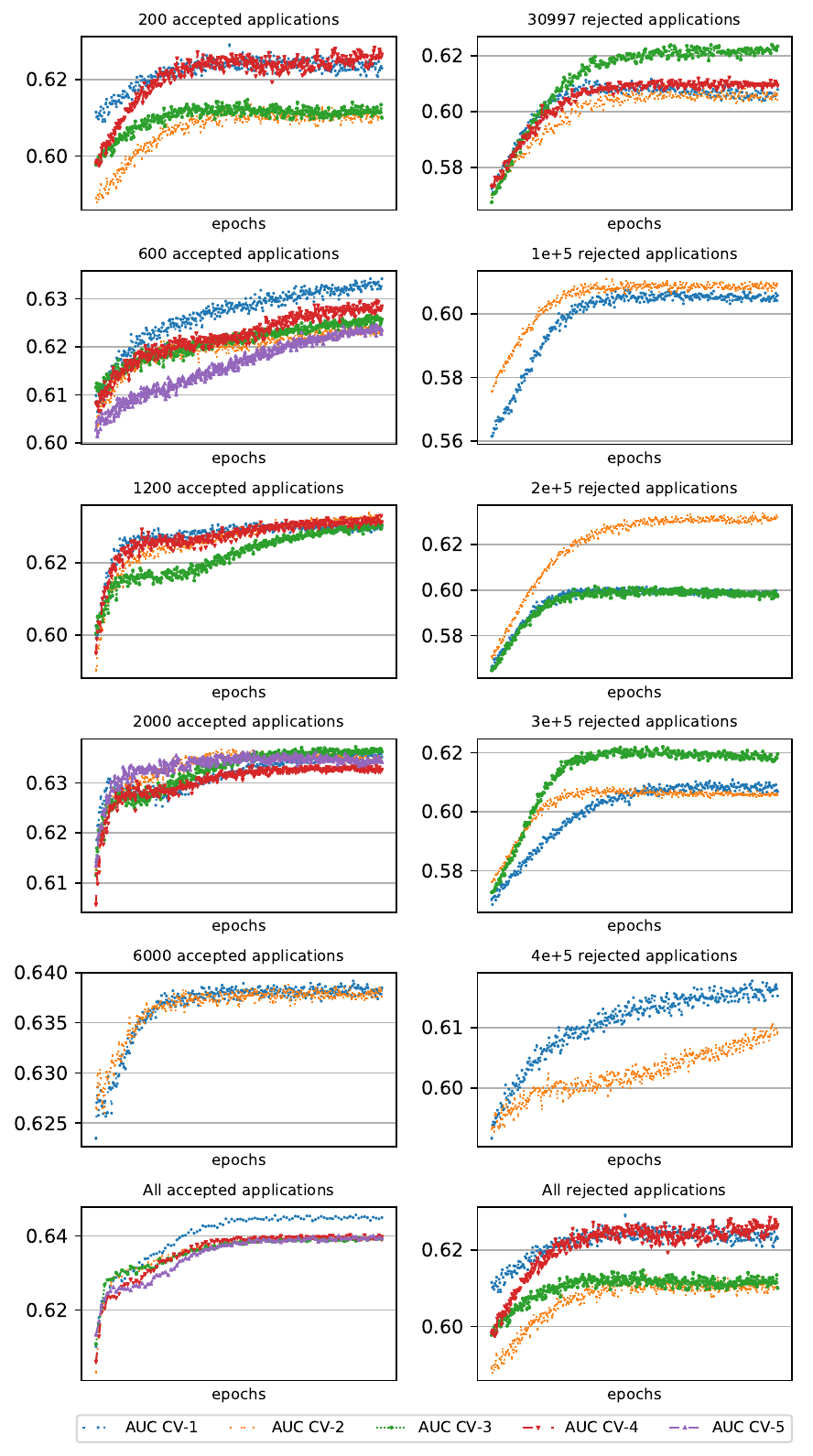}
    \caption{Model performance based on 5 cross-validations (CV) for the different scenarios analyzed in Table \ref{tbl_results4}, using the Lending Club data set and Model 2. Since training for these scenarios in some cases become unstable, we keep only the results where Model 2 converged. Note that Model 2 achieves the highest AUC equal to 0.6450 in the \textit{All} scenario in the left panel.}
    \label{fig_auc_exp1}
\end{figure}

\clearpage
\subsection{Deriving the lower bounds}\label{sec_deriving_lowerbounds}
\subsubsection{Lemma 1}
Given two multivariate Gaussian distribution, with diagonal covariance matrix, $p(\bm{x}) \sim \mathcal{N}(\bm{\mu}_1,\bm{\sigma}_1^2\bm{I})$ and $q(\bm{x}) \sim \mathcal{N}(\bm{\mu}_2,\bm{\sigma}_2^2\bm{I})$, where $\bm{\mu}_{\cdot} \in \textbf{R}^d$  and $\bm{\sigma}_{\cdot}^2 \in \textbf{R}^d$, we have:
\begin{equation}
    \int q(\bm{x}) \ \log p(\bm{x})d\bm{x} = \sum_{i=1}^d -\frac{1}{2}\log  (2\pi\sigma_{1,i}^2)-\frac{\sigma_{2,i}^2}{2\sigma_{1,i}^2}-\frac{(\mu_{2,i}-\mu_{1,i})^2}{2\sigma_{1,i}^2},
\end{equation}
where $\mu_{\cdot,i}$ and $\sigma_{\cdot,i}$ are the \textit{i}'th element of $\bm{\mu}$ and $\bm{\sigma}^2$ respectively.

\textbf{Proof:}

\begin{align}
        \int& q(\bm{x}) \ \log p(\bm{x})d\bm{x} = \int q(\bm{x})\log  \frac{1}{(2\pi)^{d/2}|\Sigma|^{1/2}}\exp{\Big(-\frac{1}{2}(\bm{x}-\bm{\mu}_1)^T \Sigma^{-1}(\bm{x}-\bm{\mu}_1)\Big)}d\bm{x} \nonumber \\
        =&-\frac{1}{2}\log  (2\pi \sigma_{1,i}^2) - \int q(\bm{x})\frac{(x_i-\mu_{1,i})^2}{2\sigma_{1,i}^2}d\bm{x}-\cdots
        -\frac{1}{2}\log  (2\pi \sigma_{1,d}^2) - \int q(\bm{x})\frac{(x_d-\mu_{1,d})^2}{2\sigma_{1,d}^2}d\bm{x} \nonumber \\
        =&- \frac{1}{2}\log (2\pi \sigma_{1,i}^2) - \frac{ \mathbb{E}_{q}[x_i^2] - 2 \mathbb{E}_{q}[x_i]\mu_{1,i}  + \mu_{1,i}^2 }{2\sigma_{1,i}^2} - \cdots - \frac{1}{2}\log (2\pi \sigma_{1,d}^2) - \frac{\mathbb{E}_{q}[x_d^2] - 2 \mathbb{E}_{q}[x_d]\mu_{1,d}  + \mu_{1,d}^2 }{2\sigma_{1,d}^2} \nonumber \\
        =&-\frac{1}{2}\log (2\pi \sigma_{1,i}^2) - \frac{\sigma_{2,i}^2+\mu_{2,i}^2 - 2 \mu_{2,i}\mu_{1,i}  + \mu_{1,i}^2 }{2\sigma_{1,i}^2} - \cdots 
        -  \frac{1}{2}\log (2\pi \sigma_{1,d}^2) - \frac{\sigma_{2,d}^2+\mu_{2,d}^2 - 2 \mu_{2,d}\mu_{1,d}  + \mu_{1,d}^2 }{2\sigma_{1,d}^2} \nonumber \\
        =&-\frac{1}{2}\log (2\pi \sigma_{1,i}^2) - \frac{\sigma_{2,i}^2+(\mu_{2,i} - \mu_{1,i})^2 }{2\sigma_{1,i}^2} - \cdots 
        -  \frac{1}{2}\log (2\pi \sigma_{1,d}^2) - \frac{\sigma_{2,d}^2+(\mu_{2,d} - \mu_{1,d})^2}{2\sigma_{1,d}^2} \nonumber \\
        =& \sum_j^d -\frac{1}{2}\log (2\pi \sigma_{1,j}^2) - \frac{\sigma_{2,j}^2}{2\sigma_{1,j}^2}-\frac{(\mu_{2,j}- \mu_{1,j})^2 }{2\sigma_{1,j}^2}. 
\end{align}
In the following sections we derive the lower bounds presented in the main text by taking the corresponding expectations, and using Lemma 1 where it is needed. We drop the subscripts $\bm{\theta}$ and $\bm{\phi}$ from the distributions $p_{\cdot}(\cdot)$ and $q_{\cdot}(\cdot)$, respectively, to do not clutter the notation. However, we use these subscripts in the parameters $\mu_{\cdot}$ and $\sigma_{\cdot}$ to distinguish between them.

\subsubsection{Model 1: Supervised lower bound}\label{deriv_m1_sup}
\begin{align*}
    \mathbb{E}_{q(\bm{z}|\bm{x},y)} [\log p(y)] =& \int q(\bm{z}|\bm{x},y) \log p(y) d\bm{z} \\
    =& \log \bm{\pi}  \\[3.5ex]
    \mathbb{E}_{q(\bm{z}|\bm{x},y)} [\log p(\bm{z}|y)] =& \int q(\bm{z}|\bm{x},y)\log p(\bm{z}|y) d\bm{z} \\
    =& \int \mathcal{N}(\bm{z};\bm{\mu}_{\bm{\phi}},\bm{\sigma}_{\bm{\phi}}^2) \log \mathcal{N}(\bm{z};\bm{\mu}_{\bm{\theta}},\bm{\sigma}_{\bm{\theta}}^2)d\bm{z} \\
    =&- \sum_{j=1}^{\ell_z}\bigg( \frac{1}{2} \log  (2\pi\sigma_{\bm{\theta}_{j,k}}^2) + \frac{\sigma_{\bm{\phi}_j}^2}{\sigma_{\bm{\theta}_{j,k}}^2} + \frac{(\mu_{\bm{\phi} _j}-\mu_{\bm{\theta}_{j,k}})^2}{\sigma_{\bm{\theta}_{j,k}}^2}\bigg) \\[3.5ex]
\end{align*}
\begin{align*}
    \mathbb{E}_{q(\bm{z}|\bm{x},y)} [\log p(\bm{x}|\bm{z})] =& \int q(\bm{z}|\bm{x},y) \log p(\bm{x}|\bm{z}) d\bm{z} \\
    \approx& \frac{1}{L} \sum_{l=1}^L \log \mathcal{N}(x_i| z_{i,l}) \\[3.5ex]
    \mathbb{E}_{q(\bm{z}|\bm{x},y)} [\log q(\bm{z}|\bm{x},y)] =& \int q(\bm{z}|\bm{x},y)\log q(\bm{z}|\bm{x},y) d\bm{z} \\
    =& \int \mathcal{N}(\bm{z};\bm{\mu}_{\bm{\phi}},\bm{\sigma}_{\bm{\phi}}^2) \log \mathcal{N}(\bm{z};\bm{\mu}_{\bm{\phi}},\bm{\sigma}_{\bm{\phi}}^2)d\bm{z} \\
    =&- \sum_{j=1}^{\ell_z}\big( \frac{1}{2} \log  (2\pi\sigma_{\bm{\phi}{j}}^2) + 1\big) 
\end{align*}

\subsubsection{Model 1: Unsupervised lower bound}\label{deriv_m1_unsup}
\begin{align*}
    \mathbb{E}_{q(\bm{z},y|\bm{x})} [\log p(y)] =& \sum_y \int q(y|\bm{x})q(\bm{z}|\bm{x},y) \log p(y) d\bm{z} \\
    =& \log \bm{\pi}  \\[3.5ex]
    \mathbb{E}_{q(\bm{z},y|\bm{x})} [\log p(\bm{z}|y)] =& \sum_y \int q(y|\bm{x})q(\bm{z}|\bm{x},y) \log p(\bm{z}|y) d\bm{z} \\
    =&\sum_y \bm{\pi}_{y|\bm{x}} \int \mathcal{N}(\bm{z};\bm{\mu}_{\bm{\phi}},\bm{\sigma}_{\bm{\phi}}^2) \log \mathcal{N}(\bm{z};\bm{\mu}_{\bm{\theta}},\bm{\sigma}_{\bm{\theta}}^2)d\bm{z} \\
    =&-\sum_y \bm{\pi}_{y|\bm{x}}\Bigg[ \sum_{j=1}^{\ell_z}\Big( \frac{1}{2} \log  (2\pi\sigma_{\bm{\theta}_{j,k}}^2) + \frac{\sigma_{\bm{\phi}_j}^2}{\sigma_{\bm{\theta}_{j,k}}^2} + \frac{(\mu_{\bm{\phi} _j}-\mu_{\bm{\theta}_{j,k}})^2}{\sigma_{\bm{\theta}_{j,k}}^2}\Big)\Bigg] \\[3.5ex]
    \mathbb{E}_{q(\bm{z},y|\bm{x})} [\log p(\bm{x}|\bm{z})] =& \sum_y \int q(y|\bm{x})q(\bm{z}|\bm{x},y) \log p(\bm{x}|\bm{z}) d\bm{z} \\
    \approx& \frac{1}{L} \sum_{l=1}^L \log \mathcal{N}(x_i| z_{i,l}) \\[3.5ex]
    \mathbb{E}_{q(\bm{z},y|\bm{x})} [\log q(\bm{z}|\bm{x},y)] =& \sum_y \int q(y|\bm{x})q(\bm{z}|\bm{x},y) \log q(\bm{z}|\bm{x},y) d\bm{z} \\
    =& \sum_y \bm{\pi}_{y|\bm{x}} \int \mathcal{N}(\bm{z};\bm{\mu}_{\bm{\phi}},\bm{\sigma}_{\bm{\phi}}^2) \log \mathcal{N}(\bm{z};\bm{\mu}_{\bm{\phi}},\bm{\sigma}_{\bm{\phi}}^2)d\bm{z} \\
    =&- \sum_y \bm{\pi}_{y|\bm{x}} \sum_{j=1}^{\ell_z}\big( \frac{1}{2} \log  (2\pi\sigma_{\bm{\phi}{j}}^2) + 1\big) 
\end{align*}
\begin{align*}
    \mathbb{E}_{q(\bm{z},y|\bm{x})} [\log q(y|\bm{x})] =& \sum_y \int q(y|\bm{x})q(\bm{z}|\bm{x},y) \log q(y|\bm{x}) d\bm{z} \\
    =& \sum_y q(y|\bm{x}) \log q(y|\bm{x})
\end{align*}

\subsubsection{Model 2: Supervised lower bound}\label{deriv_m2_sup}
\begin{align*}
     \mathbb{E}_{q(\bm{z},\bm{a}|\bm{x},y)} [\log p(y)] =& \int \int q(\bm{a}|\bm{x})q(\bm{z}|\bm{x},y)\log p(y) d\bm{z}d\bm{a} \\
     =& \log \bm{\pi} \\[3.5ex]
     \mathbb{E}_{q(\bm{z},\bm{a}|\bm{x},y)} [\log p(\bm{z}|y)]=& \int \int q(\bm{a}|\bm{x})q(\bm{z}|\bm{x},y) \log p(\bm{z}|y)d\bm{z}d\bm{a} \\
     =& \int \mathcal{N}(\bm{z};\bm{\mu}_{\bm{\phi}},\bm{\sigma}_{\bm{\phi}}^2) \log \mathcal{N}(\bm{z};\bm{\mu}_{\bm{\theta}},\bm{\sigma}_{\bm{\theta}}^2)d\bm{z} \\
    =&- \sum_{j=1}^{\ell_z}\bigg( \frac{1}{2} \log  (2\pi\sigma_{\bm{\theta}_{j,k}}^2) + \frac{\sigma_{\bm{\phi}_j}^2}{\sigma_{\bm{\theta}_{j,k}}^2} + \frac{(\mu_{\bm{\phi} _j}-\mu_{\bm{\theta}_{j,k}})^2}{\sigma_{\bm{\theta}_{j,k}}^2}\bigg) \\[3.5ex]
     \mathbb{E}_{q(\bm{z},\bm{a}|\bm{x},y)} [\log p(\bm{x}|\bm{z},y)]=& \int \int q(\bm{a}|\bm{x})q(\bm{z}|\bm{x},y) \log p(\bm{x}|\bm{z},y) d\bm{z}d\bm{a} \\
     \approx& \frac{1}{L} \sum_{l=1}^L \log \mathcal{N}(x_i| z_{i,l},y_i) \\[3.5ex]
     \mathbb{E}_{q(\bm{z},\bm{a}|\bm{x},y)} [\log p(\bm{a}) - \log q(\bm{a}|\bm{x})]=& \int \int q(\bm{a}|\bm{x})q(\bm{z}|\bm{x},y)[\log p(\bm{a}) - \log q(\bm{a}|\bm{x})]d\bm{z}d\bm{a} \\
     =& \int q(\bm{a}|\bm{x})\log p(\bm{a})d\bm{a} - \int q(\bm{a}|\bm{x})\log q(\bm{a}|\bm{x})d\bm{a} \\
     =& - \frac{1}{2} \sum_{c=1}^{\ell_a}(\sigma_{\bm{\phi}_{\bm{a}_c}}^2 + \mu_{\bm{\phi}_{\bm{a}_c}}^2 - (1+\log \sigma_{\bm{\phi}_{\bm{a}_c}}^2)) \\[3.5ex]
     \mathbb{E}_{q(\bm{z},\bm{a}|\bm{x},y)} [\log q(\bm{z}|\bm{x},y)]=& \int \int q(\bm{a}|\bm{x})q(\bm{z}|\bm{x},y)\log q(\bm{z}|\bm{x},y) d\bm{z}d\bm{a} \\
     =& \int q(\bm{z}|\bm{x},y) \log q(\bm{z}|\bm{x},y) d\bm{z} \\
     =& \frac{1}{2}\sum_{j=1}^{\ell_z}(1+\log \sigma_{\bm{\phi}_{\bm{z}_j}}^2)
\end{align*}

\subsubsection{Model 2: Unsupervised lower bound}\label{deriv_m2_unsup}
\begin{align*}
     \mathbb{E}_{q(\bm{z},\bm{a},y|\bm{x})} [\log p(y)] =& \int \sum_y \int q(\bm{a}|\bm{x})q(y|\bm{x},\bm{a})q(\bm{z}|\bm{x},y)\log p(y) d\bm{z}d\bm{a} \\
     =& \log \bm{\pi} \\[3.5ex]
     \mathbb{E}_{q(\bm{z},\bm{a},y|\bm{x})} [\log q(y|\bm{x},\bm{a})]=& \int \sum_y \int q(\bm{a}|\bm{x})q(y|\bm{x},\bm{a}) q(\bm{z}|\bm{x},y)\log q(y|\bm{x},\bm{a}) d\bm{z}d\bm{a} \\
     \approx& \frac{1}{L_a}\sum_{l_a=1}^{L_a} \sum_y q(y|\bm{x},\bm{a}_{l_a}) \log q(y|\bm{x},\bm{a}_{l_a}) 
\end{align*}

\begin{align*}
     \mathbb{E}_{q(\bm{z},\bm{a},y|\bm{x})} [\log p(\bm{z}|y)]=& \int \sum_y \int q(\bm{a}|\bm{x})q(y|\bm{x},\bm{a})q(\bm{z}|\bm{x},y) \log p(\bm{z}|y)d\bm{z}d\bm{a} \\
     \approx& \frac{1}{L_a}\sum_{l_a=1}^{L_a} \sum_y q(y|\bm{x},\bm{a}_{l_a}) \int q(\bm{z}|\bm{x},y_{l_a}) \log p(\bm{z}|y_{l_a}) d\bm{z} \\  
    \approx& - \frac{1}{L_a}\sum_{l_a=1}^{L_a}\sum_y \bm{\pi}_{y|\bm{x},a_{l_a}} \bigg[\sum_{j=1}^{\ell_z}\bigg( \frac{1}{2} \log  (2\pi\sigma_{\bm{\theta}_{j,k}}^2) + \frac{\sigma_{\bm{\phi}_j}^2}{\sigma_{\bm{\theta}_{j,k}}^2} \\
     +& \frac{(\mu_{\bm{\phi} _j}-\mu_{\bm{\theta}_{j,k}})^2}{\sigma_{\theta_{j,k}}^2}\bigg)\bigg] \\[3.5ex]
     \mathbb{E}_{q(\bm{z},\bm{a},y|\bm{x})} [\log p(\bm{x}|\bm{z},y)]=& \int \sum_y \int q(\bm{a}|\bm{x})q(y|\bm{x},\bm{a})q(\bm{z}|\bm{x},y) \log p(\bm{x}|\bm{z},y) d\bm{z}d\bm{a} \\
     \approx& \frac{1}{L_a}\sum_{l_a=1}^{L_a} \sum_y \bm{\pi}_{y|\bm{x},\bm{a}_{l_a}} \frac{1}{L_z}\sum_{l_z=1}^{L_z} \log \mathcal{N}(\bm{x}_i|\bm{z}_{i,l},y_{l_a}) \\[3.5ex]
     \mathbb{E}_{q(\bm{z},\bm{a},y|\bm{x})} [\log p(\bm{a}) - \log q(\bm{a}|\bm{x})]=& \int \sum_y \int q(\bm{a}|\bm{x})q(y|\bm{x},\bm{a})q(\bm{z}|\bm{x},y)[\log p(\bm{a}) - \log q(\bm{a}|\bm{x})]d\bm{z}d\bm{a} \\
     =& \sum_y q(\bm{a}|\bm{x}) \bigg[\int q(y|\bm{x},\bm{a})\log p(\bm{a})d\bm{a} - \int q(\bm{a}|\bm{x})\log q(\bm{a}|\bm{x})d\bm{a}\bigg] \\
     =& - \frac{1}{2} \sum_y \bm{\pi}_{y|\bm{x},a_{l_a}} \Big[\sum_{c=1}^{\ell_a}(\sigma_{\bm{\phi}_{\bm{a}_c}}^2 + \mu_{\bm{\phi}_{\bm{a}_c}}^2 - \big(1+\log \sigma_{\bm{\phi}_{\bm{a}_c}}^2)\big) \Big] \\[3.5ex]
     \mathbb{E}_{q(\bm{z},\bm{a},y|\bm{x})} [\log q(\bm{z}|\bm{x},y)]=& \int \sum_y \int q(\bm{a}|\bm{x})q(y|\bm{x},\bm{a}) q(\bm{z}|\bm{x},y)\log q(\bm{z}|\bm{x},y) d\bm{z}d\bm{a} \\
     \approx& \frac{1}{L_a}\sum_{l_a=1}^{L_a} \sum_y q(y|\bm{x},\bm{a}_{l_a})\int q(\bm{z}|\bm{x},y)\log q(\bm{z}|\bm{x},y) d\bm{z} \\
     =& - \frac{1}{L_a}\sum_{l_a=1}^{L_a} \sum_y \bm{\pi}_{y,\bm{a}_{l_a}} \Big[ \frac{1}{2}\sum_{j=1}^{\ell_z}(1+\log \sigma_{\bm{\phi}_{\bm{z}_j}}^2)\Big] \\[3.5ex]
\end{align*}

\clearpage
\bibliographystyle{unsrtnat}
\bibliography{bibliography}

\end{document}